\documentclass[12pt]{article}
 \usepackage{graphics}
 \usepackage{graphicx}

\usepackage{amssymb}
\usepackage{latexsym}
\newtheorem{Theorem}{Theorem}[section]

\newtheorem{Lemma}[Theorem]{Lemma}
\newtheorem{Corollary}[Theorem]{Corollary}
\newtheorem{Definition}[Theorem]{Definition}

\newcommand{\uline}{\vrule height.06ex depth.02ex width.6em}
\newcommand{\mvee}{\vee\kern-.69em\uline}

\begin{document}
\font\1=cmss12

\title{Standard Logics Are Valuation-Nonmonotonic}
\author{Mladen Pavi\v ci\'c\\
{\em Physics Chair, Faculty of Civil Engineering}\\ 
{\em University of Zagreb, Zagreb, Croatia}\\
\texttt{pavicic@grad.hr};
\ \texttt{http://m3k.grad.hr/pavicic}\medskip\\
Norman D.\ Megill\\
{\em Boston Information Group, 19 Locke Ln.}\\ 
{\em Lexington, MA 02420, USA}\\
\texttt{nm@alum.mit.edu};
\ \texttt{http://www.metamath.org}}
\maketitle
\begin{abstract}
It has recently been discovered that both quantum and classical
propositional logics can be modelled by classes of non-orthomodular
and thus non-distributive lattices that properly contain standard
orthomodular and Boolean classes, respectively. In this paper we
prove that these logics are complete even for those classes
of the former lattices from which the standard orthomodular
lattices and Boolean algebras are excluded. We also show that
neither quantum nor classical computers can be founded on the
latter models. It follows that logics are {\em valuation-nonmonotonic}
in the sense that their possible models (corresponding to their
possible hardware implementations) and the valuations for them
drastically change when we add new conditions to their defining
conditions. These valuations can even be completely separated by
putting them into disjoint lattice classes by a technique presented
in the paper.

\end{abstract}

\section{Introduction}
\label{sec:intro}
A good deal of artificial intelligence research is focused on
artificial neural networks, on the one hand, and on
default/nonmonotonic logic, on the other. Neural networks
are characterized by heavy reliance on logic gates. On the
other hand, nonmonotonic inference rules formalize
generalizations of standard logic that admit changes
in the sense that values of propositions may change when new
information (axioms) is added to or old information is deleted
from the system. In this paper, we show that already standard
logics (classical as well as quantum)---whose
monotonicity is usually taken for granted---are nonmonotonic at
both the level of logic gates that implement them and the level
of its valuations, i.e., mappings from the logic to its models.

We consider two standard logics (in contrast to, e.g., modal logics)
in this paper: propositional classical logic and propositional
quantum logic. In practice, classical logic relies almost
exclusively on the \{0,1\} valuation, i.e., the two-valued
truth table valuation, for its propositional
part. This valuation extends to the sentences of all theories that
make use of classical logic, such as set theory, model theory,
and the foundations of mathematics. However, there are also
non-standard valuations generated by non-distributive lattices,
which correctly model classical propositional logic, and by
non-orthomodular lattices, which correctly model quantum logic.
An immediate consequence of this valuation dichotomy
is that classical logic modelled by such non-distributive lattices
does not underlie present-day classical computers,
since non-standard valuations cannot be used to run them.
Only classical logic modelled by a Boolean algebra and having
a \{0,1\}\ valuation can serve us for such a purpose. Hence, whenever
we want to utilize a logic for a particular application we have to
specify the model we would use as well.

Before we go into details in the next sections, we should be more
specific about our distinction of standard vs.\ non-standard valuations.
Let us illustrate it with a graphical representation of the O6
lattice given in Figure \ref{fig:O6}, which can serve as a model
for classical logic in the same way that \{0,1\} Boolean algebra
can. Lines in the figure mean ordering. Thus we have
$\textstyle{0}\le x \le
y\le \textstyle{1}$ and $\textstyle{0}\le y'
\le x'\le \textstyle{1}$, where $\textstyle{0}$ and
$\textstyle{1}$ are the least and the greatest elements
of the lattice, respectively. Can this model be given a
linearly ordered or numerical interpretation,
for instance the interpretation provided by the
probabilistic semantics for classical logic \cite{leblanc-book}?
The answer is no, because when
$x\ne y\ne0,1$, an ordering between $x$ and either
$x'$ or $y'$ and between $y$ and either $x'$ or $y'$ is not defined,
and it is assumed that it cannot be defined. Hence, symbols
$\textstyle{1}$ and $\textstyle{0}$ in the figure
cannot be interpreted as the {\em numbers} $1$ and $0$.
If they were numbers,
$0<x<y<1$ and $0<y'<x'<1$ would imply that $x,y$ and $x',y'$ were
also numbers and we would, for example have $x=0.3$ and $x'=0.7$.
This means we would have  $x<x'$ and it yields $x\cap x<x'\cap x=0$,
i.e., $x=0$, which is a contradiction, since $x\ne 0$.

Therefore when we speak of {\em standard valuation} of
propositions of classical logic, we mean
any valuation for which we can establish a correspondence
with real numbers and their ordering, i.e., whose
corresponding model can be totally ordered. For instance, with
two-valued (\{{\1TRUE,FALSE}\}) Boolean algebra we can ascribe
the {\em number} 1 to {\1TRUE} and the {\em number}\/ 0 to {\1FALSE},
and in the probabilistic interpretation of classical logic
\cite{leblanc-book} all values from the interval [0,1] are
real numbers which are totally ordered.
When we deal with values from our O6 example above, there is no
way to establish a correspondence of O6 elements
with real numbers, and we shall call such a
valuation {\em non-standard}. The point here is
that the latter valuation cannot be implemented in present-day
binary computers---whose hardware usually deals with
numerical values such as voltage---and consequently also not in
the corresponding artificial intelligence, at the level of the
underlying logic gates building their hardware.

This means that a statement from a logic can be ``true'' or ``false''
in one model in one way and in some other model in another way.
When it ``holds'' (i.e., is ``true'')
in a standard model, say the two-valued Boolean algebra,
we can ascribe a number to it, say ``1''. When it ``holds'' in a
non-standard model, meaning, e.g., that it is equal to
$\textstyle{1}$ in Figure \ref{fig:O6},
we cannot do so and we cannot evaluate
the model for the statement directly with binary logic gates.

It is usually taken for granted that logic is
about propositions and their values. For example, we are
tempted to assume that proposition $p$ meaning ``Material
point $q$ is at position $\mathbf r$ at time $t$'' is either
{\em true} or {\em false}. However, with non-standard valuations
$x$ and $y$ from Figure \ref{fig:O6}, we can ascribe neither a
truth value nor even a probability to $p$, although ``$p$ or
non-$p$'' is certainly always valid meaning $p\cup p'=\textstyle{1}$.
The \{0,1\} Boolean algebra and the probabilistic model,
on the other hand, are the only known classical logical models that
allow ascribing \{0,1\} standard (i.e., numerical) values
to propositions and hence ``found[ing] the mathematical
theories of logic and probabilities''~\cite{boole}.
Classical logic defined by nothing but its axiomatic syntax is a more
general theory, in terms of the possible valuations it may have, than
its non-isomorphic semantics (e.g., a predicate logical
calculus with standard valuation\footnote{``A quantificational
schema is valid if it comes out true under all interpretations
in all nonempty universes\dots [T]he truth
value of a compound statement depends on no features of the
component sentences and terms except their truth values and
their extensions\dots [Quantificational] schema [containing sentence
letters] will be valid, clearly, just in case it resolves to
`$\top$' or to a valid schema under each substitution of `${\top}$'
and `$\bot$' for its sentence letters. So [its] test is truth-value
analysis.''\cite[p.\ 131]{quine}} which is nothing
but a ``predicate Boolean algebra'').

The standard-non-standard dichotomy can be even better understood
with the example of quantum logic which---when taken together with
its orthomodular lattice model---underlies Hilbert space and
therefore could be implemented into would-be quantum
computers and eventually into quantum artificial intelligence.
According to the Kochen-Specker theorem, a $\{0,1\}$ valuation
for quantum logic does not exist,\footnote{ In 2004 we gave
exhaustive algorithms for generation of Kochen-Specker vector
systems with arbitrary number of vectors in Hilbert spaces of
arbitrary dimension.~\cite{pmmm04b,pmmm03a,pavicic-book-05}
The algorithms use MMP (McKay-Megill-Pavi\v ci\'c) diagrams
for which in 3-dim Hilbert space a direct correspondence to
Greechie and Hasse diagrams can be established. Thus, we
also have a constructive proof of the non-existence of
a $\{0,1\}$ valuation within the lattice itself.}
but there is an analogy between a Boolean algebra
(distributive ortholattice) and an orthomodular (ortho)lattice
that underlies the Hilbert space of quantum mechanics.
Every orthomodular lattice is a model of quantum logic
just as every Boolean algebra (distributive ortholattice) is
a model of classical logic. However, as with classical logic,
there are also non-orthomodular lattices which are
models of quantum logic but on which no Hilbert space
can be built. Therefore quantum logic in general (not 
modelled by any model, i.e., without any semantics), or 
more precisely its syntax, would be of limited use
if we wanted to implement it into quantum computers.
Only one of its models---an orthomodular lattice---can
serve us for this goal, and therefore we call valuations
defined on the elements of the latter model---{\em
standard valuations}, as opposed to {\em non-standard
valuations} on the former non-orthomodular models.

In this paper, we prove the nonmonotonicity of both classical
and quantum logic with respect to particular intrinsically
different, disjoint classes of models.
The result separates two kinds of models that have so far been
assumed to belong to overlapping classes. In particular, general
families of non-distributive and non-orthomodular lattices
called weakly orthomodular and weakly distributive ortholattices
(WOML and WDOL) that are models of quantum and classical
propositional logics, respectively, for which we previously
proved soundness and completeness \cite{mpcommp99,pm-ql-l-hql1},
do include their standard models, orthomodular lattices (OML)
and Boolean algebras (BA) [distributive ortholattices (DOL)].
Here we prove that these lattices can be separated in the sense
that the logics can also be modelled by WOML and
WDOL from which the standard orthomodular and
Boolean algebras are {\em excluded}.\footnote{The
names  weakly orthomodular and weakly distributive ortholattices
stem from the fact that in general these lattice families contain
orthomodular and  distributive ones, although in the light
of the present ``disjointness results'' the names seem to
be somewhat inappropriate. Recall also that at the beginning
orthomodular lattices were called {\em weakly modular}
lattices.\ \cite{mittelstaedt-book}} Soundness and
completeness of these propositional logics are proved.

Specifically, we consider the {\em proper} subclasses of these lattice
families that exclude those lattices that are orthomodular
(for the WOML case) and distributive (for the WDOL case), i.e.,
WOML$\setminus$OML and WDOL$\setminus$BA (where ``$\setminus$'' denotes
set-theoretical difference).  Using them as the basis for a
modification of the standard Lindenbaum algebra technique, we present a
new result showing that quantum and classical propositional logics are
respectively complete for these proper subclasses, in and of themselves,
as models.  In other words, even after removing every lattice from WOML
(WDOL) in which the orthomodular (distributive) law holds, quantum
(classical) propositional logic is still complete for the remaining
lattices.

In both classical and quantum logics, when we add new conditions
to the defining conditions of the lattices that model
the logics, we get new lattices that also model these logics but
with changed valuations for the propositions from the logics.
This property of standard logics and valuations of their 
propositions is what we call {\em valuation-nonmonotonicity}. 
The more conditions we add, the fewer choices we have for 
valuations. This is why we consider
subclasses that exclude lattices obtained by adding new
conditions. For instance, WOML$\setminus$OML will provide us
only with valuations on weakly orthomodular lattices that are not
orthomodular, and by adding the orthomodularity condition to WOML
we get OML, which contains only valuations on orthomodular
lattices. Apart from the orthomodularity condition, there are many
more (if not infinitely many) conditions in between WOML and OML
that all provide different valuations and new proper subclasses,
as we show and discuss in Sections \ref{sec:nonmonoton} and
\ref{sec:compl-smaller} below.

We will study the quantum logic case first, since the results we obtain
for WOMLs will automatically hold for WDOLs and simplify our subsequent
presentation of the latter. In Section \ref{sec:newwoml}, we
define orthomodular and weakly orthomodular (ortho)lattices, and in
Section \ref{sec:newwdol} distributive and weakly distributive ones.
In Section \ref{sec:newwomloml}, we define the classes of proper weakly
orthomodular and proper weakly distributive ortholattices.
In Section \ref{sec:logic}, we define quantum and classical logics
and prove their soundness for the models defined in Section
\ref{sec:newwomloml}.  In Sections \ref{sec:compl-ql} and
\ref{sec:compl-cl}, we prove the completeness of quantum 
logic for WOML$\setminus$OML and WDOL$\setminus$BA models 
respectively. In Section \ref{sec:nonmonoton}, we define 
valuation-nonmonotonicity, and in Sections 
\ref{sec:nonmonoton} and \ref{sec:compl-smaller}, we discuss 
the differences between the completeness proofs for 
WOML$\setminus$OML, WDOL$\setminus$BA,  WOMLi$\setminus$OML, 
WDOLi$\setminus$BA, WOML$\setminus$WOMLi, and 
WDOL$\setminus$WDOLi we obtain in Sections 
\ref{sec:compl-ql}-\ref{sec:compl-smaller} and the 
completeness proofs for WOML and WDOL we obtained in 
\cite{mpcommp99,pm-ql-l-hql1}. And finally, we discuss and 
summarize the results we obtained in this paper
in Section \ref{sec:concl}.

\bigskip\bigskip

\section{Orthomodular and
Weakly Orthomodular  Lattices}
\label{sec:newwoml}

\begin{Definition}\label{def:ourOL}
An {\em ortholattice}, {\rm OL\/}, is an algebra
$\langle{\mathcal{OL}}_0,',\cup,\cap\rangle$
such that the following conditions are satisfied for any
$a,b,c\in \,{\mathcal{OL}}_0$ {\rm \cite{mpqo02}}:
\begin{eqnarray}
&&a\cup b\>=\>b\cup a\label{eq:aub}\\
&&(a\cup b)\cup c\>=\>a\cup (b\cup c)\\
&&a''\>=\>a\label{eq:notnot}\\
&&a\cup (b\cup b\,')\>=\>b\cup b\,'\\
&&a\cup (a\cap b)\>=\>a\\
&&a\cap b\>=\>(a'\cup b\,')'\label{eq:aAb}
\end{eqnarray}
In addition, since $a\cup a'=b\cup b\,'$ for any $a,b\in
\,{\mathcal{OL}}_0$, we define the {\em greatest element of
the lattice} {\em ($\textstyle{1}$)} and the {\em least element of the
lattice} \em{($\textstyle{0}$)}:
\begin{eqnarray}
\qquad\textstyle{1}\,{\buildrel\rm def\over=}a\cup a',\qquad\qquad
\rm\textstyle{0}\,{\buildrel\rm def\over =}a\cap a'\label{D:onezero}
\end{eqnarray}
and the {\rm ordering relation ($\le$) on the lattice}:
\begin{eqnarray}
\qquad a\le b\ \quad{\buildrel\rm def\over\Longleftrightarrow}\quad\ a\cap b=a
\quad\Longleftrightarrow\quad a\cup b=b
\end{eqnarray}
\end{Definition}

 Connectives  $\to_1$ ({\em Sasaki hook}),
$\to_2$ ({\em Dishkant implication}),
$\to_5$ ({\em relevance implication}),
$\to_0$ ({\em classical implication}), $\equiv $ ({\em quantum
equivalence}), and $\equiv_0$ ({\em classical equivalence})
are defined as follows:

\begin{Definition}\label{def:impl-L}
\quad $a\to_1 b\ \ {\buildrel\rm def\over =}\ \
a'\cup(a\cap b),
\quad a\to_2 b\ \ {\buildrel\rm def\over =}\ \
b'\to_1 a',\\
{}\quad\qquad a\to_5 b\ \ {\buildrel\rm def\over =}\ \
(a\cap b)\cup(a'\cap b)\cup(a'\cap b'),
\quad a\to_0 b\ \ {\buildrel\rm def\over =}\ \
a'\cup b$.
\end{Definition}

\begin{Definition}\label{L:id-bi-L}$\!\!\!$\footnote{In every
orthomodular lattice $a\equiv b=(a\to_1 b)\cap(b\to_1 a)$,
but not in every ortholattice.}
\qquad $a\equiv b\ \
{\buildrel\rm def\over =}\ \ (a\cap b)\cup(a'\cap b\,')$.
\end{Definition}

\begin{Definition}\label{L:id-bi-C}
\quad\qquad $a\equiv_0 b\ \
{\buildrel\rm def\over =}\ \ (a\to_0 b)\cap(b\to_0 a)$.
\end{Definition}

Connectives bind from weakest to strongest in the order $\to_1$ ($\to_0$),
$\equiv $ ($\equiv_0$), $\cup$, $\cap$, and $'$.

\begin{Definition}\label{def:comm}
If, in an ortholattice, $a=(a\cap b)\cup(a\cap b')$, we say
that $a$ commutes with $b$, which we write as $aCb$.
\end{Definition}

\begin{Definition}\label{def:wcomm}
If, in an ortholattice,
$a\equiv((a\cap b)\cup(a\cap b'))=\textstyle{1}$,
we say that $a$ weakly commutes with $b$, and we write
this as $aC_w b$.
\end{Definition}

\begin{Definition}\label{def:wcommr}
The commutator of $a$ and $b$, $C(a,b)$, is defined as
$(a\cap b)\cup(a\cap b')\cup(a'\cap b)\cup(a'\cap b')$.
\end{Definition}

\begin{Definition}\label{def:woml2} {\rm (Pavi\v ci\'c and Megill
 \cite{mpcommp99})}
An ortholattice in which the following condition holds:
\begin{eqnarray}
(a'\cap(a\cup b))\cup b'\cup(a\cap b)=\textstyle{1}\label{eq:woml2a}
\end{eqnarray} 
is called a {\em weakly orthomodular ortholattice} {\rm (WOML)}.
\end{Definition}

Using Definition \ref{def:impl-L}, we can also express
Eq.~(\ref{eq:woml2a}) as either of the two following equations,
which are equivalent in an ortholattice:
\begin{eqnarray}
&(a\to_2 b)'\cup(a\to_1 b)=\textstyle{1}\label{eq:woml2b}\\
&(a\to_1 b)'\cup(a\to_2 b)=\textstyle{1}\label{eq:woml2c}.
\end{eqnarray}

\begin{Definition}\label{def:oml2}
An ortholattice in which either of the following conditions
hold: {\rm\cite{mpijtp03b}}
\begin{eqnarray}
a\equiv b=\textstyle{1}
\quad\Rightarrow\quad a=b\label{eq:equiv1}\\
a\cup(a'\cap(a\cup b))=a\cup b\label{eq:oml2a}
\end{eqnarray}
 is called an {\em orthomodular lattice}
{\rm (OML)}.
\end{Definition}

The equations of Definition \ref{def:ourOL} determine a (proper) class
of lattices, called an {\em equational variety}, \cite[p.~352]{kalmb83}
that we designate OL.  Thus the term OL will have two meanings,
depending on context.  When we say a lattice is an OL, we mean that the
equations of Definition \ref{def:ourOL} hold in that lattice.  When we
say a lattice is in OL, we mean that it belongs to the equational
variety OL determined by those equations.  While these two statements
are of course equivalent, the distinction will matter when we say such
things as ``the class OL properly includes the class OML.''  Similar
remarks apply to OML, WOML, and the other varieties in this paper.

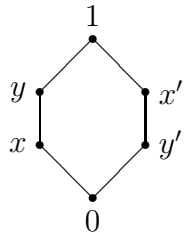
\begin{figure}[htbp]\centering
  \begin{picture}(60,80)(-10,-10)

    \put(20,0){\line(-1,1){20}}
    \put(20,0){\line(1,1){20}}
    \put(0,20){\line(0,1){20}}
    \put(40,20){\line(0,1){20}}
    \put(0,40){\line(1,1){20}}
    \put(40,40){\line(-1,1){20}}

    \put(20,-5){\makebox(0,0)[t]{$\textstyle{0}$}}
    \put(-5,20){\makebox(0,0)[r]{$x$}}
    \put(45,20){\makebox(0,0)[l]{$y'$}}
    \put(-5,40){\makebox(0,0)[r]{$y$}}
    \put(45,40){\makebox(0,0)[l]{$x'$}}
    \put(20,65){\makebox(0,0)[b]{$\textstyle{1}$}}

    \put(20,0){\circle*{3}}
    \put(0,20){\circle*{3}}
    \put(40,20){\circle*{3}}
    \put(0,40){\circle*{3}}
    \put(40,40){\circle*{3}}
    \put(20,60){\circle*{3}}

  \end{picture}
\caption{Ortholattice O6\label{fig:O6}, also called {\em benzene ring}
and {\em hexagon}.}
\end{figure}

We recall that whereas every OML is a WOML, there are WOMLs that
are not OMLs. \cite{mpcommp99}  In particular, the lattice
O6 (Fig.~\ref{fig:O6}) is a WOML but is not an OML.

On the one hand, the equations that hold in OML properly include those
that hold in WOML, since WOML is a strictly more general class of
lattices.  But there is also a sense in which the equations of WOML can
be considered to properly include those of OML, via a mapping that
Theorem \ref{th:oml-sim} below describes.  First, we need a technical lemma.

\begin{Lemma}\label{lem:womlpieces}
The following conditions hold in all {\rm WOML}s:
\begin{eqnarray}
&&a\equiv a=\textstyle{1} \label{eq:eqid}\\
&&a\equiv b=\textstyle{1}\quad\Rightarrow\quad
b\equiv a=\textstyle{1}\label{eq:eqcom} \\
&&a\equiv b=\textstyle{1}\quad\Rightarrow\quad
a'\equiv b'=\textstyle{1}\label{eq:eqnot} \\
&&a\equiv b=\textstyle{1}\quad\Rightarrow\quad (a\cup c)\equiv
        (b\cup c)=\textstyle{1}\label{eq:eqcup} \\
&&a\equiv b=\textstyle{1}\quad\Rightarrow\quad (a\cap c)\equiv
         (b\cap c)=\textstyle{1}\label{eq:eqcap} \\
&&a\equiv b=\textstyle{1}\quad\&\quad b\equiv c=
\textstyle{1}\quad\Rightarrow\quad a\equiv c=\textstyle{1}\\
&&(a\cup b)\equiv(b\cup a)=\textstyle{1}\label{eq:aubsim}\\
&&((a\cup b)\cup c)\equiv(a\cup (b\cup c))=\textstyle{1}\\
&&a''\equiv a=\textstyle{1}\label{eq:notnotsim}\\
&&(a\cup (b\cup b\,'))\equiv(b\cup b\,')=\textstyle{1}\\
&&(a\cup (a\cap b))\equiv a=\textstyle{1}\\
&&(a\cap b)\equiv(a'\cup b\,')'=\textstyle{1}\label{eq:aAbsim}\\
&&(a\cup(a'\cap(a\cup b)))\equiv(a\cup b)=
\textstyle{1}\label{eq:oml2asim}\\
&&a\equiv((a\cap b)\cup(a\cap b'))=a\equiv_0((a\cap b)\cup(a\cap b'))
  \label{eq:comcom0}\\
&&a=\textstyle{1}\quad\Leftrightarrow\quad a\equiv
\textstyle{1}=\textstyle{1}\label{eq:eq1} \\
&&a=\textstyle{1}\quad\Leftrightarrow\quad a\equiv_0
\textstyle{1}=\textstyle{1}\label{eq:eq01}
\end{eqnarray} 
In addition, Eqs.~(\ref{eq:eqid})--(\ref{eq:eqnot}) and
(\ref{eq:aubsim})--(\ref{eq:eq01}) hold in all ortholattices.
\end{Lemma}

{\parindent=0pt{\em Proof. }
Most of these conditions are proved in \cite{mpcommp99},
and the others are straightforward.
$\phantom .$\hfill$\blacksquare$}

\begin{Theorem}\label{th:oml-sim}
The equational theory of {\rm OML}s can be simulated
by a proper subset of the equational theory of {\rm WOML}s.
\end{Theorem}

{\parindent=0pt{\em Proof. }
The equational theory of OML consists of equality
axioms ($a=a$, $a=b\Rightarrow b=a$, $a=b\Rightarrow a'=b'$,
$a=b\Rightarrow a\cup c=b\cup c$,
$a=b\Rightarrow a\cap c=b\cap c$,
and $a=b\ \&\ b=c\Rightarrow a=c$); the OL
axioms, Eqs.~(\ref{eq:aub})--(\ref{eq:aAb}); and the OML law,
Eq.~(\ref{eq:equiv1}).
Any theorem of the equational variety of OMLs can be proved
with a sequence of applications of these axioms.
We construct a mapping from these axioms into
equations that hold in WOMLs as follows.
We map each axiom, which is an equation in the
form $t=s$ or an inference of the form $t_1=t_2\ldots\Rightarrow t=s$
(where $t$, $s$, and $t_1,t_2,\ldots$ are
terms), to the equation $t\equiv s=\textstyle{1}$ or the inference
$t_1\equiv t_2=\textstyle{1}\ldots\Rightarrow
t\equiv s=\textstyle{1}$.
These mappings
hold in any WOML by
Eqs.~(\ref{eq:eqid})--(\ref{eq:oml2asim}),
respectively, of Lemma \ref{lem:womlpieces}.
We then simulate the OML proof by
replacing each axiom reference
in the proof with its corresponding WOML mapping.
The result will be a proof that holds
in the equational variety of WOMLs.

Such a mapped proof will use only a proper subset of the equations
that hold in WOML: any equation whose right-hand side
does not equal $\textstyle{1}$, such as $a=a$, will never be used.
$\phantom .$\hfill$\blacksquare$}

\begin{Theorem}\label{le:teq1ded}
Let $t_1,...,t_n,t$ be any terms ($n\ge 0$).  If the inference
$t_1=\textstyle{1}\ \&\ \ldots$ $\&\ t_n=\textstyle{1}\
\Rightarrow\ t=\textstyle{1}$ holds
in all {\em OML}s, then it holds in any {\em WOML}.
\end{Theorem}

{\parindent=0pt{\em Proof. }
In any ortholattice, $t=\textstyle{1}$ iff $t\equiv
\textstyle{1}=\textstyle{1}$ by
Eq.~(\ref{eq:eq1}).  Therefore, the inference
of the theorem can be restated as follows:
$t_1\equiv \textstyle{1}=\textstyle{1}\ \&\
\ldots \&\ t_n\equiv \textstyle{1}=\textstyle{1}\
\Rightarrow\ t\equiv \textstyle{1}=\textstyle{1}$.
But this is exactly what we prove when
we simulate the original OML proof of the inference in WOML,
using the method in the proof of Theorem \ref{th:oml-sim}.  Thus
by Theorem \ref{th:oml-sim}, the inference
holds in WOML.
$\phantom .$\hfill$\blacksquare$}

\begin{Corollary}\label{cor:w-t}
No set of equations of the form $t=\textstyle{1}$ that hold
in {\rm OML}, when added to the equations of an ortholattice,
determines the equational theory of {\rm OML}s.
\end{Corollary}

{\parindent=0pt{\em Proof. }
Theorem \ref{le:teq1ded} shows that all equations of this
form hold in a WOML.
$\phantom .$\hfill$\blacksquare$}

\begin{Lemma}\label{lem:comcom}
In any {\rm WOML}, $aC_w b$ iff $C(a,b)=\textstyle{1}$.
\end{Lemma}

{\parindent=0pt{\em Proof.  }
In any OML, $aCb$ implies $a'Cb$.  Therefore, by Theorem
\ref{th:oml-sim}, $aC_w b$ implies $a'C_w b$ in any WOML.
Using Eqs.~(\ref{eq:eqcap}) and (\ref{eq:aubsim}) to combine
these two conditions, we obtain
$(a\cup a')\equiv(((a\cap b)\cup(a\cap b')) \cup((a'\cap b)\cup(a'\cap
b')))=\textstyle{1}$ i.e.,
$C(a,b)\equiv \textstyle{1}=\textstyle{1}$,
from which we obtain $C(a,b)=\textstyle{1}$
by Eq.~(\ref{eq:eq1}).
Conversely, if $C(a,b)=\textstyle{1}$, then in any OL,
$\textstyle{1}=(a\cap b)\cup(a\cap
b')\cup(a'\cap b)\cup(a'\cap b')\le (a\cap b)\cup(a\cap b')\cup a'=
(a\cap((a\cap b)\cup(a\cap b'))\cup (a'\cap((a\cap b)\cup(a\cap b'))')=
a\equiv((a\cap b)\cup(a\cap b'))$, so $aC_w b$.
$\phantom .$\hfill$\blacksquare$}

\begin{Theorem}\label{th:fh} {\rm (Foulis-Holland theorem, F-H)}
In any OML, if at least
two of the three conditions $aCb$, $aCc$, and $bCc$ hold,
then the distributive law
$a\cap(b\cup c)=(a\cap b)\cup(a\cap c)$ holds.
\end{Theorem}

{\parindent=0pt{\em Proof. }
See \cite[p.~25]{kalmb83}.
$\phantom .$\hfill$\blacksquare$}

\begin{Theorem}\label{th:wfh} {\rm (Weak Foulis-Holland theorem, wF-H)}
In any {\rm WOML}, if at least
two of the three conditions $C(a,b)=\textstyle{1}$,
$C(a,c)=\textstyle{1}$, and $C(b,c)=\textstyle{1}$ hold,
then the {\em weak distributive law}
$(a\cap(b\cup c))\equiv((a\cap b)\cup(a\cap c))=\textstyle{1}$ holds.
\end{Theorem}

{\parindent=0pt{\em Proof. }
By Lemma \ref{lem:comcom}, we can replace the conditions with
$aC_w b$, $aC_w c$, and $bC_w c$.  Then
the conclusion follows from F-H and Theorem \ref{th:oml-sim}.
$\phantom .$\hfill$\blacksquare$}

As Theorem \ref{th:oml-sim} shows, if $t$ and $s$ are terms,
then the equation $t\equiv s=\textstyle{1}$ holds in all
WOMLs iff the equation $t=s$ holds in all OMLs.
One might naively expect, then, that if $t=s$ is the OML
law, then $t\equiv s=\textstyle{1}$ will be the WOML law.
This is not always the
case:  the OML law given by Eq.~(\ref{eq:oml2a}), when converted to
$(a\cup(a'\cap(a\cup b))\equiv(a\cup b)=\textstyle{1}$,
is not the WOML law; in
fact, it holds in any OL.  However, there is a version of the OML law
with this property, as the following theorem shows.

\begin{Theorem}\label{th:oml3}
An ortholattice is an {\rm OML} iff it satisfies the following
equation:
\begin{eqnarray}
a\cup(b\cap(a'\cup b'))=a\cup b
\label{eq:oml3}
\end{eqnarray}
An ortholattice is a {\rm WOML} iff it satisfies the following
equation:
\begin{eqnarray}
(a\cup(b\cap(a'\cup b')))\equiv(a\cup b)=\textstyle{1}
\label{eq:woml3}
\end{eqnarray}
\end{Theorem}

{\parindent=0pt{\em Proof. }
For Eq.~(\ref{eq:oml3}):
It is easy to verify that Eq.~(\ref{eq:oml3}) holds in an OML,
for example by applying F-H:
$a\cup(b\cap(a'\cup b'))=
(a\cup b)\cap(a\cup a'\cup b)=(a\cup b)\cap \textstyle{1}=a\cup b$.
On the other hand, this equation fails in lattice O6
(Fig.~\ref{fig:O6}), meaning it implies the orthomodular law by
Theorem 2 of \cite[p.~22]{kalmb83}.  It is also instructive to
prove Eq.~(\ref{eq:oml2a}) directly:
$a\cup(a'\cap(a\cup b))=
a\cup((a\cup b)\cap a')=
a\cup((a\cup b)\cap (a'\cup (a'\cap b'))=
a\cup((a\cup b)\cap (a'\cup (a\cup b)'))=
a\cup(a\cup b)=
a\cup b$, where the penultimate step follows from
Eq.~(\ref{eq:oml3}) with
$a\cup b$ substituted for $b$, and all other steps hold in OL.

For Eq.~(\ref{eq:woml3}):
Since $a\cup(b\cap(a'\cup b'))=a\cup b$ holds in any OML by
Eq.~(\ref{eq:oml3}),
$(a'\cup(b'\cap(a\cup b)))\equiv(a\cup b)=\textstyle{1}$
holds in WOML by Theorem \ref{th:oml-sim}.
On the other hand, substituting $b'$ and $a'$ for $a$ and $b$ in
Eq.~(\ref{eq:woml3}), we have
$\textstyle{1}=(b'\cup(a'\cap(b''\cup a'')))\equiv(b'\cup a')
=((b'\cup(a'\cap(b\cup a)))\cap(b'\cup a'))\cup
  ((b\cap(a\cup(b'\cap a')))\cap(b\cap a))
=(b'\cup(a'\cap(b\cup a))\cup(b\cap a))
=(a'\cap(a\cup b))\cup b'\cup(a\cap b)$, which is
the WOML law Eq.~(\ref{eq:woml2a}).
$\phantom .$\hfill$\blacksquare$}

Another version of the WOML law will be useful later.

\begin{Theorem}\label{woml4}
An ortholattice is a {\rm WOML} iff it satisfies
the following condition:
\begin{eqnarray}
a\to_1 b=\textstyle{1}\qquad\Rightarrow\qquad a
\to_2 b=\textstyle{1}
\label{eq:woml4}
\end{eqnarray}
\end{Theorem}

{\parindent=0pt{\em Proof. }
See Theorem 3.9 of \cite{mpcommp99}.
$\phantom .$\hfill$\blacksquare$}

\section{Distributive and Weakly Distributive\\ Ortholattices}
\label{sec:newwdol}

\begin{Definition}\label{def:wdol} {\rm (Pavi\v ci\'c and
Megill \cite{mpcommp99})}
An ortholattice in which the following equation holds:
\begin{eqnarray}
(a\equiv b)\cup(a\equiv b')=
(a\cap b)\cup(a\cap b')\cup(a'\cap b)\cup(a'\cap b')=\textstyle{1}
\label{eq:wdol}
\end{eqnarray} 
is called a {\em weakly distributive ortholattice}, {\rm WDOL}.
\end{Definition}

A WDOL is thus an ortholattice in which the condition
$C(a,b)=\textstyle{1}$ holds.
This condition is known as {\em commensurability}.~\cite[Def.~(2.13),
p.~32]{mittelstaedt-book}.

\begin{Definition}\label{def:ba} An
ortholattice to which the following condition is added:
\begin{eqnarray}
a\cap(b\cup c)=(a\cap b)\cup(a\cap c)\label{eq:ba}
\end{eqnarray} 
is called a distributive ortholattice {\rm (DOL)} or (much more
often) a {\em Boolean algebra} {\rm (BA)}.
\end{Definition}

Eq.~(\ref{eq:ba}) is called the {\em distributive law}.

We recall that whereas every BA is a WDOL, there are WDOLs that
are not BAs. \cite{mpcommp99}  In particular, the lattice
O6 (Fig.~\ref{fig:O6}) is a WDOL but is not a BA.

The first part of the following theorem will turn out to be very useful,
because it will let us reuse all of the results we have already obtained
for WOMLs.

\begin{Theorem}\label{th:wdolwoml}
Every {\rm WDOL} is a {\rm WOML}, but
not every {\rm WOML} is a {\rm WDOL}.
\end{Theorem}

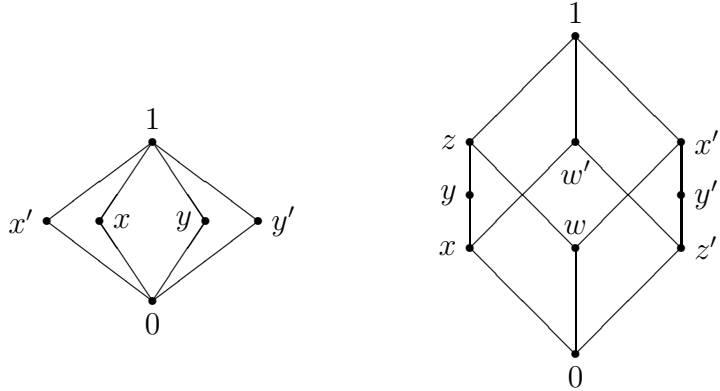
\begin{figure}[htbp]\centering
  \begin{picture}(260,150)(-10,-10)

    \put(0,20){

      \begin{picture}(90,80)(0,0)
        \put(40,0){\line(-2,3){20}}
        \put(40,0){\line(2,3){20}}
        \put(40,0){\line(-4,3){40}}
        \put(40,0){\line(4,3){40}}
        \put(40,60){\line(-2,-3){20}}
        \put(40,60){\line(2,-3){20}}
        \put(40,60){\line(-4,-3){40}}
        \put(40,60){\line(4,-3){40}}

        \put(40,-5){\makebox(0,0)[t]{$\textstyle{0}$}}
        \put(25,30){\makebox(0,0)[l]{$x$}}
        \put(85,30){\makebox(0,0)[l]{$y'$}}
        \put(55,30){\makebox(0,0)[r]{$y$}}
        \put(-5,30){\makebox(0,0)[r]{$x'$}}
        \put(40,65){\makebox(0,0)[b]{$\textstyle{1}$}}

        \put(40,0){\circle*{3}}
        \put(0,30){\circle*{3}}
        \put(20,30){\circle*{3}}
        \put(60,30){\circle*{3}}
        \put(80,30){\circle*{3}}
        \put(40,60){\circle*{3}}
      \end{picture}
    } 

    \put(150,0){
       \begin{picture}(100,125)(0,0)

    \put(50,0){\line(-1,1){40}}
    \put(50,0){\line(0,1){40}}
    \put(50,0){\line(1,1){40}}
    \put(10,40){\line(0,1){20}}
    \put(10,40){\line(1,1){40}}
    \put(50,40){\line(-1,1){40}}

    \put(50,40){\line(1,1){40}}
    \put(90,40){\line(-1,1){40}}
    \put(90,40){\line(0,1){20}}
    \put(10,60){\line(0,1){20}}
    \put(90,60){\line(0,1){20}}
    \put(10,80){\line(1,1){40}}
    \put(50,80){\line(0,1){40}}
    \put(90,80){\line(-1,1){40}}

    \put(50,-5){\makebox(0,0)[t]{$\textstyle{0}$}}
    \put(5,40){\makebox(0,0)[r]{$x$}}
    \put(50,45){\makebox(0,0)[b]{$w$}}
    \put(95,40){\makebox(0,0)[l]{$z'$}}
    \put(5,60){\makebox(0,0)[r]{$y$}}
    \put(95,60){\makebox(0,0)[l]{$y'$}}
    \put(5,80){\makebox(0,0)[r]{$z$}}
    \put(50,73){\makebox(0,0)[t]{$w'$}}
    \put(95,80){\makebox(0,0)[l]{$x'$}}
    \put(50,125){\makebox(0,0)[b]{$\textstyle{1}$}}

    \put(50,0){\circle*{3}}
    \put(10,40){\circle*{3}}
    \put(50,40){\circle*{3}}
    \put(90,40){\circle*{3}}
    \put(10,60){\circle*{3}}
    \put(90,60){\circle*{3}}
    \put(10,80){\circle*{3}}
    \put(50,80){\circle*{3}}
    \put(90,80){\circle*{3}}
    \put(50,120){\circle*{3}}

      \end{picture}
    } 

  \end{picture}
  \caption{\hbox to3mm{\hfill}(a) OML M02; \hbox to15mm{\hfill} (b)
           Non-WDOL from \protect\cite{mphpa98}, Fig.~3.
        \label{fig:nonwdol}}
\end{figure}

{\parindent=0pt{\em Proof. }
Since $a'\cap b\le a'\cap(a\cup b)$ and $(a\cap b')\cup(a'\cap b')\le
b'$ in any OL, the WDOL law, Eq.~(\ref{eq:wdol}), gives us
$\textstyle{1}=(a\cap
b)\cup(a\cap b')\cup(a'\cap b)\cup(a'\cap b')\le (a'\cap(a\cup b))\cup
b'\cup(a\cap b)$, which is the WOML law, Eq.~(\ref{eq:woml2a}).

On the other hand, the modular (and therefore WOML) lattice MO2
(Fig.~\ref{fig:nonwdol}a) violates Eq.~(\ref{eq:wdol}).  If we put $x$
for $a$ and $y$ for $b$, the equation evaluates to
$\textstyle{0}=\textstyle{1}$.
$\phantom .$\hfill$\blacksquare$}

We are now in a position to prove two important equivalents to the WDOL
law.  We call them {\em weak distributive laws}, since they provide
analogs to the distributive law of Boolean algebras.

\begin{Theorem}\label{th:wdol2}
An ortholattice is a {\rm WDOL} iff it satisfies either
of the following
equations:
\begin{eqnarray}
(a\cap(b\cup c))\equiv_0((a\cap b)\cup(a\cap c))=
\textstyle{1}\label{eq:wdol2} \\
(a\cap(b\cup c))\equiv((a\cap b)\cup(a\cap c))=
\textstyle{1}\label{eq:wdol3}
\end{eqnarray}
\end{Theorem}

{\parindent=0pt{\em Proof.  }
First, we prove these laws can be derived from each other
in any OL.

Assuming Eq.~(\ref{eq:wdol2}) and using the fact that $(a\cap
b)\cup(a\cap c)\le (a\cap(b\cup c)$, in any OL we have
$\textstyle{1}=((a\cap(b\cup
c))\to_0((a\cap b)\cup(a\cap c)) )\cap( ((a\cap b)\cup(a\cap
c))\to_0(a\cap(b\cup c)) )=((a\cap(b\cup c))\to_0((a\cap b)\cup(a\cap
c))$.  Putting $b'$ for $c$,
$\textstyle{1}=((a\cap(b\cup b'))\to_0((a\cap
b)\cup(a\cap b'))= (a\to_0((a\cap b)\cup(a\cap b'))=(a'\cup((a\cap
b)\cup(a\cap b'))\le (b'\cap (b\cup a))\cup a'\cup(b\cap a)$, which is
the WOML law.  This lets us use our previous WOML results.

Starting from the last equality in the first sentence of the previous
paragraph, in any OL we also have
$\textstyle{1}=((a\cap(b\cup c))\to_0((a\cap
b)\cup(a\cap c))=(a\cap(b\cup c))\to_1((a\cap b)\cup(a\cap
c))=((a\cap(b\cup c))\to_1((a\cap b)\cup(a\cap c)) )\cap( ((a\cap
b)\cup(a\cap c))\to_1(a\cap(b\cup c)) )$.  Therefore, using the footnote
to Definition \ref{L:id-bi-L} and Theorem \ref{le:teq1ded}, it follows
that in any WOML, and therefore (by the previous paragraph) in any OL,
Eq.~(\ref{eq:wdol2}) implies
$\textstyle{1}=(a\cap(b\cup c))\equiv ((a\cap b)\cup(a\cap c))$.

Conversely, Eq.~(\ref{eq:wdol2}) follows immediately from
Eq.~(\ref{eq:wdol3}) in any OL.  Thus these two equations are
equivalent laws when added to the equations for OL.

Next, we prove that Eq.~(\ref{eq:wdol3}) is equivalent to the
WDOL law in the presence of the equations for OL.

Since $C(a,b)=\textstyle{1}$ for any $a,b$ in a WDOL,
Eq.~(\ref{eq:wdol3}) follows
immediately from wF-H (Theorem \ref{th:wfh}), which holds in every WOML
and thus, by Theorem \ref{th:wdolwoml}, in every WDOL.

Conversely, in OML, we can prove $C(a,b)=\textstyle{1}$ if
we use instances of the
distributive law as hypotheses.  Using Theorem \ref{th:oml-sim}, such a
proof can be converted to a WOML proof, replacing the instances of the
distributive law with instances of Eq.~(\ref{eq:wdol3}).  This will
yield a proof of $C(a,b)\equiv \textstyle{1}=\textstyle{1}$,
which in any OL implies $C(a,b)=\textstyle{1}$ by
Eq.~(\ref{eq:eq1}).  This proves that Eq.~(\ref{eq:wdol3}) implies the
WDOL law, Eq.~(\ref{eq:wdol}).
$\phantom .$\hfill$\blacksquare$}

\begin{Theorem}\label{th:wdol5}
An ortholattice is a {\rm WDOL} iff it satisfies either of
the following equations:
\begin{eqnarray}
&a\equiv((a\cap b)\cup(a\cap b'))=\textstyle{1}\label{eq:wdol5}\\
&a\equiv_0((a\cap b)\cup(a\cap b'))=\textstyle{1}\label{eq:wdol6}.
\end{eqnarray}
\end{Theorem}

{\parindent=0pt{\em Proof.  }
In any OL, $a\equiv((a\cap b)\cup(a\cap b'))=
(a\cap((a\cap b)\cup(a\cap b')))\cup(a'\cap((a'\cup b')\cap(a'\cup b)))=
((a\cap b)\cup(a\cap b'))\cup a'=
(a\to_0 b)'\cup(a\to_1 b)$.  Thus Eq.~(\ref{eq:wdol5}) implies
$\textstyle{1}=(a\to_0 b)'\cup(a\to_1 b) \le
(a\to_2 b)'\cup(a\to_1 b)$, which
is the WOML law in the form of
Eq.~(\ref{eq:woml2b}).  By Lemma \ref{lem:comcom}, in any WOML
Eq.~(\ref{eq:wdol5}) implies $C(a,b)=\textstyle{1}$,
which is the WDOL law.

For the converse, Eq.~(\ref{eq:wdol5}) holds in an WDOL by
Lemma \ref{lem:comcom}.

Eq.~(\ref{eq:wdol6}) is equivalent to Eq.~(\ref{eq:wdol5}) in
any OL by Eq.~(\ref{eq:comcom0}).
$\phantom .$\hfill$\blacksquare$}

We mention that Eq.~(\ref{eq:wdol5}) is the definition of $aC_w b$.

\begin{Theorem}\label{th:wdol4}
An ortholattice is a {\rm WDOL} iff it satisfies
the following condition:
\begin{eqnarray}
a\equiv_0 b=\textstyle{1}\qquad\Rightarrow\qquad (a\cup c)\equiv_0
                 (b\cup c)=\textstyle{1}\label{eq:wdol4}
\end{eqnarray}
\end{Theorem}

{\parindent=0pt{\em Proof.  }
First, we show that Eq.~(\ref{eq:wdol4}) implies the WOML law.  Putting
$d$ for $a$ and $d\cap e$ for $b$, the hypothesis becomes, in an OL,
$\textstyle{1}=d\equiv_0
(d\cap e)=(d'\cup(d\cap e))\cap((d'\cup e')\cup d)=
(d'\cup(d\cap e))\cap \textstyle{1}=d\to_1 e$.
Also putting $e$ for $c$, the
conclusion becomes, in an OL,
$\textstyle{1}=(d\cup e)\equiv_0 ((d\cap e)\cup
e)=(d\cup e)\equiv_0 e=((d'\cap e')\cup e) \cap(e'\cup (d\cup
e))=((d'\cap e')\cup e) \cap \textstyle{1}=d\to_2 e$.
The condition $d\to_1
e=\textstyle{1}\ \Rightarrow\ d\to_2
e=\textstyle{1}$ is the WOML law by Eq.~(\ref{eq:woml4}).

Having our previous WOML results now available to us, we next show that
Eq.~(\ref{eq:wdol4}) implies the WDOL law.  We put $d'\cap(d\cup e')$
for $a$, $e'\cap(e\cup d')$ for $b$, and $d'$ for $c$.  To satisfy the
hypothesis, we must show that in any WOML, $(d'\cap(d\cup e'))\equiv_0
(e'\cap(e\cup d'))= \textstyle{1}$, i.e., that
$((d\cup(d'\cap e))\cup (e'\cap(e\cup
d')))\cap ((e\cup(e'\cap d))\cup
(d'\cap(d\cup e')))=\textstyle{1}$.  For the first
conjunct, we apply wF-H to $(d\cup(d'\cap e))\cup (e'\cap(e\cup d'))=
((d\cup(d'\cap e))\cup (e'\cap(e\cup d')))\equiv
\textstyle{1}=\textstyle{1}$ to obtain
$)(d\cup(d'\cap e)\cup e')\cap(d\cup (d'\cap e)\cup e\cup d'))\equiv
\textstyle{1}=\textstyle{1}$, which reduces to
$(\textstyle{1}\cap \textstyle{1})\equiv
\textstyle{1}=\textstyle{1}$.  The other conjunct is
satisfied similarly, by symmetry.  The conclusion becomes
$((d'\cap(d\cup e'))\cup d')\equiv_0 ((e'\cap(e\cup d'))\cup
d')=d'\equiv_0 ((e'\cap(e\cup d'))\cup d')= \textstyle{1}$.
Expanding the definition of $\equiv_0$ and discarding the
left-hand conjunct, we have
$((e\cup(e'\cap d))\cap d)\cup d'=\textstyle{1}$.
Using wF-H, this becomes
$\textstyle{1}=((e\cap d)\cup((e'\cap d)\cap d))\cup
d'= ((e\cap d)\cup(e'\cap d))\cup d'=
(((e\cap d)\cup(e'\cap d))\cup d')\equiv \textstyle{1}$.
Conjoining both sides of the $\equiv$ with $d$ using
Eq.~(\ref{eq:eqcap}), we have
$((((e\cap d)\cup(e'\cap d))\cup d')\cap d)\equiv
(\textstyle{1}\cap d)=\textstyle{1}$.
Applying wF-H twice, we obtain $\textstyle{1}=((((e\cap d)\cap
d)\cup((e'\cap d)\cap d))\cup (d'\cap d))\equiv
(\textstyle{1}\cap d)= (((e\cap d)\cup(e'\cap d))\cup
\textstyle{0})\equiv d= ((e\cap d)\cup(e'\cap d))\equiv d$,
which is the WDOL law in the form of Eq.~(\ref{eq:wdol5}).

Conversely, to show that Eq.~(\ref{eq:wdol4}) holds in any WDOL, we
apply Eq.~(\ref{eq:eqeq0}) below (which does not depend on the present
theorem) to the hypothesis and conclusion, converting it to
Eq.~(\ref{eq:eqcup}).
$\phantom .$\hfill$\blacksquare$}

An essential characteristic of the WDOL law and its equivalents is
that they must fail in the modular
(and therefore OML and WOML) lattice MO2.
However, such a failure is not sufficient to ensure that we have
a WDOL law equivalent.

\begin{Theorem}\label{th:eqeq0}
The following condition holds in all {\rm WDOL}s:
\begin{eqnarray}
a\equiv_0 b=\textstyle{1}\quad\Leftrightarrow\quad
a\equiv b=\textstyle{1}\label{eq:eqeq0}
\end{eqnarray}
It also fails in modular lattice {\rm MO2}.  However, when added to the
equations for {\em OL}, it does not determine the equations of {\rm WDOL}.
\end{Theorem}

{\parindent=0pt{\em Proof.  } To verify that this condition holds in
any WDOL, we first convert the hypothesis to the OL-equivalent
hypothesis  $(a\equiv_0 b)\equiv \textstyle{1}=\textstyle{1}$
using Eq.~(\ref{eq:eq01}).
By using the WDOL law $C(a,b)=\textstyle{1}$ to satisfy the hypotheses
of any uses of wF-H, it is then
easy to prove that this condition holds in
any WDOL.
In particular, the reverse implication holds in any OL.

The failure
of Eq.~(\ref{eq:eqeq0}) in MO2 is verified by putting $x$ for $a$ and
$y$ for $b$; then the left-hand side holds but the right-hand side
becomes $\textstyle{0}=\textstyle{1}$.
On the other hand, it does not imply the
WDOL law nor even the WOML law:  it passes in the non-WOML
lattice of Figure \ref{fig:nonwdol}b.
$\phantom .$\hfill$\blacksquare$}

On the one hand, the equations that hold in BA properly include those
that hold in WDOL, since WDOL is a strictly more general class of
lattices.  But there is also a sense in which the equations of WDOL can
be considered to properly include those of BA, via a mapping that
Theorem \ref{th:ba-sim} below describes.

\begin{Theorem}\label{th:ba-sim}
The equational theory of {\rm BA}s can be simulated
by a proper subset of the equational theory of {\rm WDOL}s.
\end{Theorem}

{\parindent=0pt{\em Proof. }
The equational theory of BA consists of equality axioms (see the proof
of Theorem \ref{th:oml-sim}); the OL axioms,
Eqs.~(\ref{eq:aub})--(\ref{eq:aAb}); and the distributive law,
Eq.~(\ref{eq:ba}).  Any theorem of the equational variety of BAs can be
proved with a sequence of applications of these axioms.  We construct a
mapping from these axioms into equations that hold in WDOLs as follows.
We map each axiom, which is an equation in the form $t=s$ or an
inference of the form $t_1=t_2\ldots\Rightarrow t=s$ (where $t$, $s$,
and $t_1,t_2,\ldots$ are terms), to the equation
$t\equiv_0 s=\textstyle{1}$ or the
inference $t_1\equiv_0 t_2=\textstyle{1}\ldots\Rightarrow
t\equiv_0 s=\textstyle{1}$.  These
mappings hold in any WDOL by Eqs.~(\ref{eq:eqid})--(\ref{eq:aAbsim}) and
(\ref{eq:wdol2}), respectively, after converting $\equiv$ to $\equiv_0$
with Eq.~(\ref{eq:eqeq0}).  We then simulate the BA proof by replacing
each axiom reference in the proof with its corresponding WDOL mapping.
The result will be a proof that holds in the equational variety of
WDOLs.

Such a mapped proof will use only a proper subset of the equations
that hold in WDOL: any equation whose right-hand side
does not equal $\textstyle{1}$, such as $a=a$, will never be used.
$\phantom .$\hfill$\blacksquare$}

\begin{Theorem}\label{le:teq1dedba}
Let $t_1,...,t_n,t$ be any terms ($n\ge 0$).
If the inference $t_1=\textstyle{1}\ \&\ \ldots$ $\&\
t_n=\textstyle{1}\ \Rightarrow\ t=\textstyle{1}$ holds
in all {\em BA}s, then it holds in any {\em WDOL}.
\end{Theorem}

{\parindent=0pt{\em Proof. }
In any ortholattice, $t=\textstyle{1}$ iff $t\equiv_0
\textstyle{1}=\textstyle{1}$ by
Eq.~(\ref{eq:eq01}).  Therefore, the inference
of the theorem can be restated as follows:
$t_1\equiv_0 \textstyle{1}=\textstyle{1}\ \&\
\ldots \&\ t_n\equiv_0 \textstyle{1}=\textstyle{1}\
\Rightarrow\ t\equiv_0 \textstyle{1}=\textstyle{1}$.
But this is exactly what we prove when
we simulate the original BA proof of the inference in WDOL,
using the method in the proof of Theorem \ref{th:ba-sim}.  Thus
by Theorem \ref{th:ba-sim}, the inference
holds in WDOL.
$\phantom .$\hfill$\blacksquare$}

\begin{Corollary}\label{cor:w-tba}
No set of equations of the form $t=\textstyle{1}$ that hold
in {\rm BA}, when added to the equations of an ortholattice,
determines the equational theory of {\rm BA}s.
\end{Corollary}

{\parindent=0pt{\em Proof. }
Theorem \ref{le:teq1dedba} shows that all equations of this
form hold in a WDOL.
$\phantom .$\hfill$\blacksquare$}

\section{The Classes of Proper Weakly\\ 
Orthomodular and Proper Weakly\\ 
Distributive Ortholattices}
\label{sec:newwomloml}

One of the main aims of our paper is to prove that both quantum
and classical logics are sound and complete with respect to at least
a class of all weakly orthomodular lattices (WOMLs) in which
orthomodularity fails for every lattice and a class of all
weakly distributive lattices
(WDOLs) in which distributivity fails for every lattice, respectively.

To prove the soundness and completeness of quantum logic we
shall consider a new class of lattices that belong to the
class WOML but not to the class OML. We will denote the
resulting class WOML-OML.  In other words, WOML-OML denotes the
set-theoretical difference $\mbox{\rm WOML}\setminus\mbox{\rm OML}$.  A
member of the class WOML-OML is a lattice, specifically a member of the
class WOML, and we will call such a lattice a {\em proper} WOML.  Thus a
proper WOML is one that satisfies the WOML equations but violates the
OML equations.  Lattice O6 is an example of a proper WOML.  Lattice MO2
is an example of a WOML that is not a proper WOML, i.e., that does not
belong to the class WOML-OML, since it also belongs to the class OML.

Notice that WOML-OML is not an equational variety like WOML, because we
cannot turn WOML into WOML-OML by adding new equational conditions to
those defining WOML.
If we try to add the orthomodularity condition (\ref{eq:equiv1})
\cite{p98,mpijtp03b} to WOML-OML, we will get the empty set.

In Section \ref{sec:compl-ql} we shall show that quantum logics is
complete for WOML-OML: every wff whose valuation equals
$\textstyle{1}$ for
all members of WOML-OML is a provable statement in quantum logic.
This is not necessarily obvious {\em a priori}: quantum logic
($\mathcal{QL}$) is not necessarily complete for an arbitrary collection
of WOMLs.  For example, it is not complete for the subset of
WOML-OML consisting of the singleton set $\{\mbox{\rm O}6\}$,
since O6 is a model for classical logic.

The significance of this result can be explained as follows.  Since
$\mathcal{QL}$ is already complete for OML models, it might be argued
that completeness for the more general WOML models (\cite{mpcommp99})
has its origin in the OML members of the equational variety WOML, rather
than being an intrinsic property of the non-OML members.  We show that
this is not the case by completely removing all OMLs from the picture.

In order for the completeness proof to go through, we will have to
construct a special Lindenbaum algebra that belongs to WOML-OML.  This
requires a modification to the standard Lindenbaum algebra (which, in
the standard proof, ``wants'' to be an OML).  The technique that we use,
involving cutting down the equivalence classes for the Lindenbaum
algebra to force it to belong to WOML-OML, might be useful for other
completeness proofs that are not amenable to the standard
Lindenbaum-algebra approach.

Following an analogous blueprint, in Section \ref{sec:compl-cl} we
will also show that classical
logic is complete for the class of models WDOL-BA, defined as the
set-theoretical difference $\mbox{\rm WDOL}\setminus\mbox{\rm BA}$
(where WDOL and BA here denote equational varieties), which again
by definition has nothing to do with Boolean algebras.  In fact,
a simpler result is possible:  Schechter \cite[p.~272]{schechter}
has proved that classical logic ($\mathcal{CL}$) is complete for
the single ${\rm WDOL}$ lattice O6.  Schechter's result can be
strengthened to show that classical logic is complete for {\em any}
subset of WDOL.  This is an immediate consequence of the fact that
classical logic is maximal, i.e., no extension of it can be consistent.
So if classical logic is sound for a model, it is automatically complete
for that model.

\section{Logics and Their Soundness for\\ Our Models}
\label{sec:logic}

Logic ($\mathcal{L}$) is a language consisting of
propositions and a set of conditions
and rules imposed on them called axioms and rules of inference.

The propositions we use are well-formed formulae (wffs),
defined as follows.
We denote elementary, or primitive, propositions by
$p_0,p_1,p_2,...$, and have the following primitive connectives:
$\neg$ (negation) and $\vee$ (disjunction).
The set of wffs is defined recursively as follows:
\begin{enumerate}
\item[] $p_j$ is a wff for $j=0,1,2,...$
\item[] $\neg A$ is a wff if $A$ is a wff.
\item[] $A\vee B$ is a wff if $A$ and $B$ are wffs.
\end{enumerate}

We introduce conjunction with the following definition:

\begin{Definition} \label{D:conj}
$A\wedge B\ {\buildrel\rm def\over =}\ \neg (\neg A\vee\neg B)$.
\end{Definition}

\medskip
The operations of implication are the following ones (classical, Sasaki,
and Kalmbach) \cite{pav87}:

\begin{Definition} \label{def:impl-0}
$\qquad A\to_0 B\ \ {\buildrel\rm def\over =} \ \
\neg A\vee B$.
\end{Definition}

\begin{Definition} \label{def:impl-1}
$\qquad A\to_1 B\ \ {\buildrel\rm def\over =} \ \
\neg A\vee (A\wedge B)$.
\end{Definition}

\begin{Definition} \label{def:impl-3}
$\qquad A\to_3 B\ \ {\buildrel\rm def\over =} \ \
(\neg A\wedge B)\vee(\neg A\wedge\neg B)\vee(A\wedge(\neg A\vee B))$.
\end{Definition}

We also define the {\em equivalence} operations as follows:

\begin{Definition}\label{L:equiv}
$\qquad A\equiv B\ \
{\buildrel\rm def\over =}\ \ (A\wedge B)\vee(\neg A\wedge\neg B)$.
\end{Definition}

\begin{Definition}\label{L:equiv-0}
$\qquad A\equiv_0 B\ \
{\buildrel\rm def\over =}\ \ (A\to_0 B)\wedge(B\to_0 A)$.
\end{Definition}

Connectives bind from weakest to strongest in the order $\to$,
$\equiv$, $\vee$, $\wedge$, $\neg$.

\smallskip
Let $\mathcal{F}^\circ$ be the set of all propositions, i.e.,
of all wffs.
Of the above connectives, $\vee$ and $\neg$ are primitive ones.
Wffs containing $\vee$ and $\neg$ within logic $\mathcal{L}$
are used to build an algebra
${\mathcal{F}}=\langle {\mathcal{F}}^\circ,\neg,\vee\rangle$.
In $\mathcal{L}$, a set of axioms and rules of inference are imposed on
${\mathcal{F}}$. {}From a set of axioms by means of rules of inference,
we get other expressions which we call theorems. Axioms themselves
are also theorems.
A special symbol $\vdash$ is used to denote the set of theorems.
Hence $A\in\ \vdash$ iff $A$ is a theorem. The statement
$A\in\ \vdash$ is usually written as $\vdash A$. We read this: ``$A$ is
provable'' since if $A$ is a theorem, then there is a proof for it.
We present the axiom systems of our propositional logics
in schemata form (so that we dispense with the rule of
substitution).

\subsection{Quantum Logic and Its Soundness for\\ WOML-OML Models}
\label{subsec:q-logic}

We present Kalmbach's quantum logic because it is
the system that has been investigated in the greatest detail
in her book \cite{kalmb83} and elsewhere \cite{kalmb74,mphpa98}.
Quantum logic ($\mathcal{QL}$) is defined as a language
consisting of propositions and connectives (operations) as introduced
above, and the following axioms and a rule of inference.
We will use $\vdash_\mathcal{QL}$ to denote provability from
the axioms and rule of $\mathcal{QL}$ and omit the subscript when
it is clear from context (such as in the list of axioms that follow).

\smallskip
\noindent{\bf Axioms}
\begin{eqnarray}
{\rm A1}\qquad &&\vdash A\label{eq:kalmb-A1}\equiv A\\
{\rm A2}\qquad &&\vdash A\equiv B\rightarrow_0(B\equiv C\rightarrow_0
     A\equiv C)\\
{\rm A3}\qquad &&\vdash A\equiv B\rightarrow_0\neg A\equiv \neg B\\
{\rm A4}\qquad &&\vdash A\equiv B\rightarrow_0A\wedge C\equiv B\wedge C\\
{\rm A5}\qquad &&\vdash A\wedge B\equiv B\wedge A\\
{\rm A6}\qquad &&\vdash A\wedge (B\wedge C)\equiv (A\wedge B)\wedge C\\
{\rm A7}\qquad &&\vdash A\wedge (A\vee B)\equiv A\\
{\rm A8}\qquad &&\vdash\neg A\wedge A\equiv(\neg A\wedge A)\wedge B \\
{\rm A9}\qquad &&\vdash A\equiv\neg\neg A\\
{\rm A10}\qquad &&\vdash\neg(A\vee B)\equiv\neg A\wedge\neg B\\
{\rm A11}\qquad &&\vdash A\vee(\neg A\wedge(A\vee B))\equiv A\vee B\label{eq:a11}\\
{\rm A12}\qquad &&\vdash (A\equiv B)\equiv(B\equiv A)\\
{\rm A13}\qquad &&\vdash A\equiv B\rightarrow_0(A\rightarrow_0 B)\\
{\rm A14}\qquad &&\vdash (A\to_0B)\to_3(A\to_3(A\to_3B))\\
{\rm A15}\qquad &&\vdash (A\to_3B)\to_0(A\to_0 B)\label{eq:kalmb-A15}
\end{eqnarray}
{\bf Rule of Inference} ({\em Modus Ponens})
\begin{eqnarray}
\ {\rm R1}\qquad &&\vdash A \quad \& \quad \vdash A \rightarrow_3 B
\quad\Rightarrow \quad \vdash B\label{eq:kalmb-R1}
\end{eqnarray}
In Kalmbach's presentation, the connectives $\vee$, $\wedge$, and $\neg$
are primitive.  In the base set of any model (such as an OML or WOML
model) that belongs to OL, $\cap$ can be defined
in terms of $\cup$ and $'$, as justified by DeMorgan's law, and
thus the corresponding $\wedge$ can be defined in terms of
$\vee$ and $\neg$ [using Eq.~(\ref{eq:aAb})].  We shall do this for
simplicity.  Regardless of whether
we consider $\wedge$ primitive or defined,
we can drop axioms A1, A11, and A15 because it has been
proved that they are redundant, i.e., can be derived from the
other axioms.~\cite{mphpa98}

\begin{Definition}\label{D:gamma-ql}
For\/ $\Gamma\subseteq{\mathcal{F}}^\circ$ we say $A$ is derivable from\/
$\Gamma$ and write\/ $\Gamma\vdash_\mathcal{QL} A$ or just\/
$\Gamma\vdash A$ if there is a finite sequence of
formulae, the last of which is $A$, and
each of which is either one of the axioms of
$\mathcal{QL}$ or is a member of\/ $\Gamma$ or is obtained from its
precursors with the help of a rule of inference of the logic.
\end{Definition}

To prove soundness means to prove that all axioms as well
as the rules of inference (and therefore all theorems) of
$\mathcal{QL}$ hold in its models.

\begin{Definition}\label{exists-c}We call ${\mathcal{M}}=\langle
L,h\rangle$ a model if $L$ is an algebra and
$h:{\mathcal{F}}^\circ\longrightarrow L$, called a valuation,
is a morphism of formulae ${\mathcal{F}}^\circ$
into $L$, preserving the operations $\neg,\vee$
while turning them into $',\cup$.
\end{Definition}

Whenever the base set $L$ of a model belongs to WOML-OML, we
say (informally) that the model belongs to WOML-OML.  In particular, if
we say ``for all models in WOML-OML'' or ``for all proper WOML models,'' we
mean for all base sets in WOML-OML and for all valuations on each base
set.  The term ``model'' may refer either to a specific pair
$\langle L,h\rangle$ or to all possible such pairs with the
base set $L$, depending on context.

\begin{Definition}\label{one}We call a formula $A\in{\mathcal{F}}^\circ$
valid in the model $\mathcal{M}$, and write $\vDash_\mathcal{M} A$, if
$h(A)=\textstyle{1}$ for all valuations $h$ on the model, i.e.,
for all $h$ associated with the base set $L$ of the model.
We call a formula $A\in{\mathcal{F}}^\circ$ a consequence of\/
$\Gamma\subseteq{\mathcal{F}}^\circ$ in the model $\mathcal{M}$ and
write $\Gamma\vDash_\mathcal{M} A$ if $h(X)=\textstyle{1}$ for
all $X$ in\/ $\Gamma$ implies $h(A)=\textstyle{1}$,
for all valuations $h$.  \end{Definition}

\medskip

For brevity, whenever we do not make it explicit, the notations
$\vDash_\mathcal{M} A$ and $\Gamma\vDash_\mathcal{M} A$ will always be
implicitly quantified over all models of the appropriate type, in this
section for all proper WOML models $\mathcal{M}$.  Similarly, when we say
``valid'' without qualification, we will mean valid in all models of
that type.

The following theorem shows that if $A$ is a theorem of
${\mathcal{QL}}$, then $A$ is valid in any proper WOML model.

In \cite{mpcommp99,pm-ql-l-hql1} we proved the soundness
for WOML and OML. We now prove the soundness of quantum logic
by means of WOML-OML, i.e., that if $A$ is a theorem in
${\mathcal{QL}}$, then $A$ is valid in any proper WOML model,
i.e., in any WOML-OML model.

\begin{Theorem}\label{th:soundness}{\rm [Soundness]}
$\qquad\Gamma\vdash A\quad\Rightarrow\quad\Gamma\vDash_\mathcal{M} A$
\end{Theorem}

{\parindent=0pt{\em Proof. }
By Theorem 29 of \cite{pm-ql-l-hql2},
any WOML is a model for ${\mathcal{QL}}$.  Therefore,
any proper WOML is also a model.
$\phantom .$\hfill$\blacksquare$}

\subsection{Classical Logic and Its Soundness for\\ WDOL-BA Models}
\label{subsec:c-logic}

We make use of the PM classical logical system $\mathcal{CL}$
(Whitehead and Russell's \it Principia Mathematica\/ \rm axiomatization
in Hilbert and Ackermann's presentation \cite{hilb-ack-book} but
in schemata form so that we dispense with their rule of
substitution).  In this system, the connectives $\vee$ and $\neg$
are primitive, and the $\to_0$ connective shown in the axioms is
implicitly understood to be expanded according to its definition.
We will use $\vdash_\mathcal{CL}$ to denote provability from
the axioms and rule of $\mathcal{CL}$, omitting the subscript when
it is clear from context.

\smallskip
\noindent{\bf Axioms}
\begin{eqnarray}
{\rm A1}\qquad &&\vdash A\vee A\to_0 A\label{eq:cl-a1}\\
{\rm A2}\qquad &&\vdash A\to_0 A\vee B\\
{\rm A3}\qquad &&\vdash A\vee B\to_0 B\vee A\\
{\rm A4}\qquad &&\vdash (A\to_0 B)\to_0(C\vee A\to_0 C\vee B)\label{eq:cl-a4}
\end{eqnarray}
{\bf Rule of Inference} ({\em Modus Ponens})
\begin{eqnarray}
{\rm R1}\qquad &&\vdash A \qquad \&\qquad
A\to_0 B\qquad\Rightarrow\qquad\vdash B\label{eq:cl-r1}
\end{eqnarray}

We assume that the only legitimate way of inferring theorems in
$\mathcal{CL}$ is by means of these axioms and the Modus
Ponens rule. We make no assumption about valuations of the
primitive propositions from which wffs are built, but instead
are interested in wffs that are valid
in the underlying models.  Soundness and completeness will
show that those theorems that can be inferred from the
axioms and the rule of inference are exactly those that are valid.

We define derivability in ${\mathcal{CL}}$,
$\Gamma\vdash_\mathcal{CL} A$ or just $\Gamma\vdash A$, in the
same way as we do for system ${\mathcal{QL}}$. The models
and validity of formulae in a model are also defined as for
${\mathcal{QL}}$ above.

The following theorem shows that if $A$ is a theorem of
${\mathcal{CL}}$, then $A$ is valid in any proper WDOL model.

In \cite{mpcommp99,pm-ql-l-hql1} we proved the soundness
for WDOL and BA. We now prove the soundness of classical logic
by means of WDOL-BA, i.e., that if $A$ is a theorem in
${\mathcal{CL}}$, then $A$ is valid in any proper WDOL model,
i.e., in any WDOL-BA model.

\begin{Theorem}\label{th:c-soundness}{\rm [Soundness]}
$\qquad\Gamma\vdash A\quad\Rightarrow\quad\Gamma\vDash_\mathcal{M} A$
\end{Theorem}

{\parindent=0pt{\em Proof. }
By Theorem 30 of \cite{pm-ql-l-hql2},
any WDOL is a model for ${\mathcal{CL}}$.  Therefore,
any proper WDOL is also a model.
$\phantom .$\hfill$\blacksquare$}

\section{The Completeness of Quantum Logic for WOML-OML
Models}
\label{sec:compl-ql}

Our main task in proving the soundness of $\mathcal{QL}$ in the previous
section was to show that all axioms as well as the rules of inference
(and therefore all theorems) from $\mathcal{QL}$ hold in WOML-OML. The task
of proving the completeness of $\mathcal{QL}$ is the opposite one:
we have to impose the structure of WOML-OML on the set
${\mathcal{F}}^\circ$ of formulae of $\mathcal{QL}$.

We start with a relation of congruence, i.e.,
a relation of equivalence compatible with the operations in
$\mathcal{QL}$. We make use of an equivalence relation to
establish a correspondence between formulae of $\mathcal{QL}$ and
formulae of WOML-OML. The resulting equivalence classes stand for elements
of a proper WOML (i.e., a member
of WOML-OML) and enable the completeness proof of $\mathcal{QL}$ by means
of WOML-OML.

Our definition of congruence involves a special set of valuations
on lattice O6 (shown in Figure \ref{fig:O6})
called ${\mathcal{O}}${\rm 6} and defined as follows.

\begin{Definition}\label{D:hexagon}Letting {\em O6} represent the lattice
from Figure \ref{fig:O6}, we define ${\mathcal{O}}${\rm 6} as the set of all
mappings $o:{\mathcal{F}}^\circ\longrightarrow {\rm O}6$ such that for
$A,B\in{\mathcal{F}}^\circ$,
$o(\neg A)=o(A)'$, and $o(A\vee B)=o(A)\cup o(B)$.
\end{Definition}

The purpose of ${\mathcal{O}}${\rm 6} is to let us refine the
equivalence classes used for the completeness proof, so that the
Lindenbaum algebra will be a proper WOML, i.e., one that is not
orthomodular.  This is accomplished by conjoining the term $(\forall
o\in{\mathcal{O}}6)[(\forall X\in\Gamma)(o(X)=\textstyle{1})
\Rightarrow o(A)=o(B)]$
to the equivalence relation definition, meaning that for equivalence we
require also that (whenever the valuations $o$ of the wffs in $\Gamma$ are
all $\textstyle{1})$ the valuations of wffs $A$ and $B$ map to
the same point in the
lattice O6. Thus wffs $A\vee B$ and $A\vee (\neg
A\wedge (A\vee B))$ become members of two separate equivalence
classes, what by Theorem \ref{th:non-distr} below, amounts to
non-orthomodularity of WOML.  Without the conjoined
term, these two wffs would belong to the same equivalence class.  The
point of doing this is to provide a completeness proof that is not
dependent in any way on the orthomodular law and to show that
completeness does not require that any of the underlying models
be OMLs.

\begin{Theorem}\label{th:congruence-nonoml}
The relation of {\em equivalence} $\approx_{\Gamma,\mathcal{QL}}$
or just $\approx$, defined as
\begin{eqnarray}&&\hskip-5ptA\approx B\\&&\hskip8pt{\buildrel\rm
def\over =}\
\Gamma\vdash
A\equiv B\ \&\ (\forall o\in{\mathcal{O}}{\rm 6})[(\forall
X\in\Gamma)(o(X)=\textstyle{1})
\Rightarrow o(A)=o(B)],\nonumber
\label{eq:equiv-noml}
\end{eqnarray}
is a relation of congruence in the algebra
$\mathcal{F}$, where\/ $\Gamma\subseteq{\mathcal{F}}^\circ$
\end{Theorem}

{\parindent=0pt{\em Proof. } Let us first prove that $\approx$ is an equivalence
relation. $\>A\approx A\>$ follows from A1
[Eq.~(\ref{eq:kalmb-A1})] of system $\mathcal{QL}$ and the identity
law of equality.
If $\Gamma\vdash A\equiv B$, we can detach the left-hand
side of A12 to conclude $\Gamma\vdash B\equiv A$, through the use of
A13 and repeated uses of A14 and R1.  From this and commutativity
of equality, we conclude $\>A\approx B\>\Rightarrow\>
B\approx A$.  (For brevity we will
not usually
mention further uses of A12, A13, A14, and R1 in what follows.)
 The proof of transitivity runs as
follows.
\begin{eqnarray}
A\approx B&&\quad\&\quad B\approx C\label{line1}\\
&&\Rightarrow\ \Gamma\vdash A\equiv   B\quad\&\quad \Gamma\vdash
B\equiv   C\nonumber\\
&&\hskip-20pt\&\ (\forall o\in{\mathcal{O}}6)
[(\forall X\in\Gamma)(o(X)=\textstyle{1})\ \Rightarrow\
o(A)=o(B)]\nonumber\\
&&\hskip-20pt\&\ (\forall o\in{\mathcal{O}}6)
[(\forall X\in\Gamma)(o(X)=\textstyle{1})\ \Rightarrow\
o(B)=o(C)]\nonumber\\
&&\Rightarrow\ \Gamma\vdash A\equiv   C \nonumber\\
&&\hskip-20pt\&\ (\forall o\in{\mathcal{O}}6)[(\forall
X\in\Gamma)(o(X)=\textstyle{1})\ \Rightarrow\
o(A)=o(B)\ \&\ o(B)=o(C)].\nonumber
\end{eqnarray}
In the last line above, $\Gamma\vdash A\equiv C$ follows from A2, and
the last metaconjunction reduces to  $\ o(A)=o(C)\ $
by transitivity of equality.
Hence the conclusion $A\approx C$ by definition.

In order to be a relation of congruence, the relation of
equivalence must be compatible with the operations $\neg$ and
$\vee$. These proofs run as follows.
\begin{eqnarray}
A\approx B&&\label{line-1}\\
&&\Rightarrow\Gamma\vdash A\equiv B\nonumber\\
&&\hskip-20pt\&\ \ (\forall o\in{\mathcal{O}}6)
[(\forall X\in\Gamma)(o(X)=\textstyle{1})\
\Rightarrow\ o(A)=o(B)]\nonumber\\
&&\Rightarrow\Gamma\vdash\neg A\equiv\neg B\nonumber\\
&&\hskip-20pt\&\ \ (\forall o\in{\mathcal{O}}6)
[(\forall X\in\Gamma)(o(X)=\textstyle{1})\
\Rightarrow\ o(A)'=o(B)']
\nonumber\\
&&\Rightarrow\Gamma\vdash\neg A\equiv\neg B\nonumber\\
&&\hskip-20pt\&\ \ (\forall o\in{\mathcal{O}}6)
[(\forall X\in\Gamma)(o(X)=\textstyle{1})\
\Rightarrow\ o(\neg A)=o(\neg B)]
\nonumber\\
&&\Rightarrow\neg A\approx\neg B\nonumber
\end{eqnarray}
\begin{eqnarray}
A\approx B&&\label{line-11}\\
&&\Rightarrow \Gamma\vdash A\equiv B\nonumber\\
&&\hskip-20pt\&\ \ (\forall o\in{\mathcal{O}}6)
[(\forall X\in\Gamma)(o(X)=\textstyle{1})\
\Rightarrow\ o(A)=o(B)]\nonumber\\
&&\Rightarrow\Gamma\vdash(A\vee C)\equiv(B\vee C)\nonumber\\
&&\hskip-30pt\&\ \ (\forall o\in{\mathcal{O}}6)
[(\forall X\in\Gamma)(o(X)=\textstyle{1})\
\Rightarrow\ o(A)\cup o(C)=o(B)\cup o(C)]
\nonumber\\
&&\Rightarrow(A\vee C)\approx(B\vee C)\nonumber
\end{eqnarray}
In the second step of Eq.~(\ref{line-1}), we used A3.  In the
second step of Eq.~(\ref{line-11}), we used A4 and A10.
For the quantified part of these expressions, we applied
the definition of ${\mathcal{O}}6$.
$\phantom .$\hfill$\blacksquare$}

\begin{Definition}\label{D:equiv-class-sets-woml}
The equivalence class for wff $A$ under the relation of equivalence
$\approx$ is
defined as $|A|=\{B\in {\mathcal{F}}^\circ:A\approx B\}$, and we denote
${\mathcal{F}}^\circ/\!\approx\ =\{|A|:A\in {\mathcal{F}}^\circ\}$.
The equivalence classes define the natural morphism
$f:{\mathcal{F}}^\circ\longrightarrow
{\mathcal{F}}^\circ/\!\approx$, which gives
$f(A)\ =^{\rm def}\ |A|$. We write $a=f(A)$, $b=f(B)$, etc.
\end{Definition}

\begin{Lemma}\label{L:equality-non-q}
The relation $a=b$ on ${\mathcal{F}}^\circ/\!\approx$ is given by:
\begin{eqnarray}
\hskip80pt |A|=|B|\qquad&\Leftrightarrow&\qquad A\approx B
\label{eq:equation-non-om-q}
\end{eqnarray}
\end{Lemma}

\begin{Lemma}\label{L:lind-alg-non-q} The Lindenbaum algebra
${\mathcal{A}}=\langle {\mathcal{F}}^\circ/\!\approx,\neg/\!\approx,
\vee/\!\approx\rangle$ is a
{\rm WOML}, i.e., Eqs.~(\ref{eq:aub})--(\ref{eq:aAb})
 and Eq.~(\ref{eq:woml2a}) hold for
$\neg/\!\approx$ and $\vee/\!\approx$
as  $'$ and $\cup$
respectively {\em [where---for simplicity---we use the same symbols
($'$ and $\cup$) as for O6, since there are no ambiguous
expressions in which the origin of the operations would not
be clear from the context]}.
\end{Lemma}

{\parindent=0pt{\em Proof. }
For the $\Gamma\vdash A\equiv B$ part of the $A\approx B$ definition,
the proofs of the ortholattice conditions,
Eqs.~(\ref{eq:aub})--(\ref{eq:aAb}), follow from
A5, A6, A9, the dual of A8, the dual of A7, and DeMorgan's laws
respectively.
(The duals follow from DeMorgan's laws, derived from A10, A9, and A3.)
A11 gives us an analog of the OML law for the $\Gamma\vdash A\equiv B$
part, and the WOML law
Eq.~(\ref{eq:woml2a}) follows from the OML law in an ortholattice.
For the quantified part of the $A\approx B$ definition,
lattice O6 is a (proper) WOML.
$\phantom .$\hfill$\blacksquare$}

\begin{Lemma}\label{L:lind-alg-eq-1-q}
In the Lindenbaum algebra $\mathcal{A}$, if
$f(X)=\textstyle{1}$ for all $X$ in\/ $\Gamma$ implies
$f(A)=\textstyle{1}$,
then\/ $\Gamma\vdash A$.
\end{Lemma}

{\parindent=0pt{\em Proof. }
Let us assume that $f(X)=\textstyle{1}$ for all $X$ in
$\Gamma$ implies $f(A)=\textstyle{1}$ i.e.,
$|A|=\textstyle{1}=|A|\cup |A|'=|A\vee\neg A|$, where the
first equality is from
Definition \ref{D:equiv-class-sets-woml}, the
second equality follows from Eq.~(\ref{D:onezero})
(the definition of $\textstyle{1}$ in an ortholattice), and
the third from the fact that $\approx$ is a congruence.
Thus
$A \approx (A\vee\neg A)$, which by definition means
$\Gamma\vdash
A\equiv (A\vee\neg A)\ \&\ (\forall o\in{\mathcal{O}}6)[(\forall
X\in\Gamma)(o(X)=\textstyle{1})
\Rightarrow o(A)=o((A\vee\neg A))]$. This implies, in particular
(by dropping the second conjunct), $\Gamma\vdash
A\equiv (A\vee\neg A)$.  Now in any ortholattice, $a\equiv (a\cup a')=a$
holds.  By mapping the steps in the proof of this ortholattice
identity to steps in a proof in the logic, we can prove
$\vdash (A\equiv (A\vee\neg A))\equiv A$ from $\mathcal{QL}$
axioms A1--A15.  (We call this a ``proof by analogy,'' which is
closely related to the method of Theorem \ref{th:oml-sim}.  A direct proof
of $\vdash (A\equiv (A\vee\neg A))\equiv A$ is also not difficult.)
Detaching the left-hand side (using A12, A13, A14, and R1), we
conclude $\Gamma\vdash A$.
$\phantom .$\hfill$\blacksquare$}

\begin{Theorem}\label{th:non-distr}The orthomodular law does not hold
in $\mathcal{A}$.
\end{Theorem}

{\parindent=0pt{\em Proof. } This is Theorem 3.27 from \cite{mpcommp99}, and
the proof provided there runs as follows.
We assume ${\mathcal F}^\circ$ contains at least two elementary
(primitive) propositions $p_0,p_1,\ldots$. We pick
a valuation $o$ that maps two of them, $A$ and $B$, to
distinct nodes $o(A)$ and $o(B)$ of O6 that are
neither $\textstyle{0}$ nor $\textstyle{1}$ such
that $o(A)\le o(B)$
[i.e., $o(A)$ and $o(B)$ are on the same side of hexagon O6 in
Figure \ref{fig:O6}].
From the structure of O6, we obtain
$\>o(A)\cup o(B)=o(B)$ and $o(A)\cup(o(A)'\cap(o(A)\cup o(B)))
=o(A)\cup(o(A)'\cap o(B))= o(A)\cup \textstyle{0}=o(A)$.
Therefore $o(A)\cup o(B)\ne o(A)\cup (o(A)' \cap (o(A)\cup o(B))$,
i.e., $o(A\vee B)\ne o(A\vee (\neg A\wedge(A\vee B)))$.
This falsifies $(A\vee B)\approx (A\vee(\neg A\wedge (A\vee B))$.
Therefore $a\cup b\ne a\cup (a'\cap(a\cup b))$,
providing a counterexample to the orthomodular law
for ${\mathcal F}^\circ/\!\approx$.
$\phantom .$\hfill$\blacksquare$}

\begin{Lemma}\label{L:model-wdol}$\mathcal{M}_\mathcal{A}=\langle
\mathcal{A},f\rangle$ is a proper {\rm WOML} model.
\end{Lemma}

{\parindent=0pt{\em Proof. } Follows from Lemma \ref{L:lind-alg-non-q} and
Theorem \ref{th:non-distr}. $\phantom .$\hfill$\blacksquare$}

Now we are able to prove the completeness of $\mathcal{QL}$, i.e.,
that if a formula {\rm A} is a consequence of a set
of wffs $\Gamma$
in all {\rm WOML-OML} models,
then $\Gamma\vdash A$.  In particular, when $\Gamma=\varnothing$,
all valid formulae are provable in $\mathcal{QL}$.  (Recall from
the note below Definition \ref{one} that
the left-hand side of the metaimplication below is implicitly
quantified over all proper WOML models $\mathcal{M}$.)

\begin{Theorem}\label{th:completeness-woml}{\rm[Completeness]}
$\qquad\Gamma\vDash_\mathcal{M} A\qquad\Rightarrow\qquad
\Gamma\vdash A$.
\end{Theorem}

{\parindent=0pt{\em Proof. }
$\Gamma\vDash_\mathcal{M} A$ means that in all proper WOML models
$\mathcal{M}$, if $f(X)=\textstyle{1}$ for all $X$ in $\Gamma$,
then $f(A)=\textstyle{1}$ holds.   In particular, it holds for
$\mathcal{M}_\mathcal{A}=\langle
\mathcal{A},f\rangle$, which is a proper WOML model
by Lemma \ref{L:model-wdol}.  Therefore, in the
Lindenbaum algebra $\mathcal{A}$, if
$f(X)=\textstyle{1}$ for all $X$ in $\Gamma$, then
$f(A)=\textstyle{1}$ holds.
By Lemma \ref{L:lind-alg-eq-1-q}, it follows that
$\Gamma\vdash A$.
$\phantom .$\hfill$\blacksquare$}

\section{The Completeness of Classical Logic for\\ WDOL-BA
Models}
\label{sec:compl-cl}

We have to impose the structure of WDOL-BA on the set
${\mathcal{F}}^\circ$ of formulae of $\mathcal{CL}$.
We start with a relation of congruence, i.e.,
a relation of equivalence compatible with the operations in
$\mathcal{CL}$. We make use of an equivalence relation to
establish a correspondence between formulae of $\mathcal{QL}$ and
formulae of WDOL-BA. The resulting equivalence classes stand for
elements of a proper WDOL (i.e., a member of WDOL-BA) and enable
the completeness proof of $\mathcal{QL}$ by means
of WDOL-BA. We will closely follow the procedure outlined
in Section \ref{sec:compl-ql} and will often implicitly assume
that definitions and theorems given in that section for
 $\mathcal{QL}$ have a completely analogous form for
$\mathcal{CL}$.

\begin{Theorem}\label{th:congruence-nondist}
The relation of {\em equivalence} $\approx_{\Gamma,\mathcal{CL}}$
or just $\approx$, defined as
\begin{eqnarray}&&\hskip-5ptA\approx B\\&&\hskip8pt{\buildrel\rm
def\over =}\
\Gamma\vdash
A\equiv_0 B\ \&\ (\forall o\in{\mathcal{O}}6)[(\forall
X\in\Gamma)(o(X)=\textstyle{1})
\Rightarrow o(A)=o(B)],\nonumber
\label{eq:equiv-noml-c}
\end{eqnarray}
is a relation of congruence in the algebra
$\mathcal{F}$.
\end{Theorem}

{\parindent=0pt{\em Proof. }
As given in \cite{pm-ql-l-hql2}.
$\phantom .$\hfill$\blacksquare$}

\begin{Lemma}\label{L:lind-alg-non-q1} The Lindenbaum algebra
${\mathcal{A}}=\langle {\mathcal{F}}^\circ/\!\approx,\neg/\!\approx,
\vee/\!\approx\rangle$ is a
{\rm WDOL}, i.e., Eqs.~(\ref{eq:cl-a1})--(\ref{eq:cl-a4})
 and Eq.~(\ref{eq:cl-r1}) hold for
$\neg/\!\approx$ and $\vee/\!\approx$
as  $'$ and $\cup$
respectively.
\end{Lemma}

{\parindent=0pt{\em Proof. }In analogy to Lemma \ref{L:lind-alg-non-q} and
following \cite{pm-ql-l-hql2}.
$\phantom .$\hfill$\blacksquare$}

\begin{Lemma}\label{L:lind-alg-eq-1-c}
In the Lindenbaum algebra $\mathcal{A}$, if
$f(X)=\textstyle{1}$ for all $X$ in\/ $\Gamma$ implies
$f(A)=\textstyle{1}$,
then\/ $\Gamma\vdash A$.
\end{Lemma}

{\parindent=0pt{\em Proof.}
As given in \cite{pm-ql-l-hql2}.
$\phantom .$\hfill$\blacksquare$}

\begin{Theorem}\label{th:non-distr-c}Distributivity does not hold
in $\mathcal{A}$.
\end{Theorem}

{\parindent=0pt{\em Proof.}
$(a\cap(b\cup c))=((a\cap b)\cup(a\cap c))$ fails in O6.
$\phantom .$\hfill$\blacksquare$}

\begin{Lemma}\label{L:model-wdol-c}$\mathcal{M}_\mathcal{A}=\langle
\mathcal{A},f\rangle$ is a proper {\rm WDOL} model.
\end{Lemma}

{\parindent=0pt{\em Proof.}
Follows from Lemma \ref{L:lind-alg-non-q1} and
Theorem~\ref{th:non-distr-c}.
$\phantom .$\hfill$\blacksquare$}

\begin{Theorem}\label{th:completeness-wdol-c}{\rm[Completeness]}
$\qquad\Gamma\vDash_\mathcal{M} A\quad \Rightarrow\quad \Gamma\vdash A$
\end{Theorem}

{\parindent=0pt{\em Proof.}
Analogous to the proof of Theorem \ref{th:completeness-woml}.
$\phantom .$\hfill$\blacksquare$}

\section{Valuation-Nonmonotonicity}
\label{sec:nonmonoton}

In Sections \ref{sec:logic}, \ref{sec:compl-ql}, and
\ref{sec:compl-cl} we prove the soundness and completeness
of both quantum ($\mathcal{QL}$) and classical ($\mathcal{CL}$)
standard logic for proper weakly orthomodular (WOML-OML) and
weakly distributive (WDOL-BA) ortholattices, respectively.
As we stressed in the Introduction and in Section
\ref{sec:newwomloml}, WOML-OML is the class of all those
ortholattices (see Definition \ref{def:ourOL}) that satisfy
Definition \ref{def:woml2} (WOML) but do not satisfy Definition
\ref{def:oml2}. Analogously, WDOL-BA includes all
those ortholattices  that satisfy Definition \ref{th:wdol5}
but do not satisfy Definition \ref{def:ba}.

The set-theoretical differences WOML$\setminus$OML (WOML-OMLs)
and WDOL$\setminus$BA (WDOL-BAs)  determine valuations that
quantum and classical logic can respectively make use of.
The set of valuations that can be assigned to logical
propositions are simply elements of any of particular
lattices, e.g., O6 given in Figure \ref{fig:O6}.
Of course, any standard Boolean valuation set such as,
e.g., \{0,1\}, i.e., \{{\1TRUE,FALSE}\}, is then
precluded by definition. On the other hand, if we
decide to use, e.g., \{0,1\}-valuation, i.e., two-valued BA as
our model, then we cannot use WDOL-BAs valuations any more.

Both WOML-OMLs and WDOL-BAs, on the one hand, and
OMLs and BAs, on the other, are models for which we
can prove soundness and completeness of quantum and
classical logic, respectively. Which ones we will use,
i.e., which valuations we will choose, depends on the
hardware, i.e., the kind of implementation we adopt.
For  an implementation of the \{0,1\} valuation, we use
today's binary chips; for the O6 or any other non-Boolean
valuation, we might design appropriate chips and circuits
in the future. Actually there are certainly many more
non-Boolean valuations than the O6 one, if not
infinitely many.

\begin{figure}[htbp]\centering
    \setlength{\unitlength}{1pt}  
    \begin{picture}(260,150)(-10,-10)
      \put(65,0) { 
        \begin{picture}(100,100)(0,0) 
          \put(0,25){\line(0,1){75}}
          \put(0,25){\line(2,-1){50}}
          \put(0,25){\line(1,1){50}}
          \put(0,50){\line(2,-1){50}}
          \put(0,50){\line(1,1){25}}
          \put(0,100){\line(1,-1){37.5}}
          \put(0,100){\line(2,1){50}}
          \put(50,0){\line(0,1){50}}
          \put(50,125){\line(0,-1){50}}
          \put(50,50){\line(-3,1){37.5}}
          \put(100,100){\line(0,-1){75}}
          \put(100,100){\line(-2,1){50}}
          \put(100,100){\line(-1,-1){50}}
          \put(100,75){\line(-2,1){50}}
          \put(100,75){\line(-1,-1){25}}
          \put(100,25){\line(-1,1){37.5}}
          \put(100,25){\line(-2,-1){50}}
          \put(50,75){\line(3,-1){37.5}}

          \put(0,25){\circle*{3}}
          \put(0,50){\circle*{3}}
          \put(0,75){\circle*{3}}
          \put(0,100){\circle*{3}}
          \put(12.5,62.5){\circle*{3}}
          \put(25,75){\circle*{3}}
          \put(37.5,62.5){\circle*{3}}
          \put(50,0){\circle*{3}}
          \put(50,25){\circle*{3}}
          \put(50,50){\circle*{3}}
          \put(50,75){\circle*{3}}
          \put(50,100){\circle*{3}}
          \put(50,125){\circle*{3}}
          \put(62.5,62.5){\circle*{3}}
          \put(75,50){\circle*{3}}
          \put(87.5,62.5){\circle*{3}}
          \put(100,25){\circle*{3}}
          \put(100,50){\circle*{3}}
          \put(100,75){\circle*{3}}
          \put(100,100){\circle*{3}}

          \put(0,25){\makebox(0,0)[r]{$w\ $}}
          \put(0,50){\makebox(0,0)[r]{$z\ $}}
          \put(0,75){\makebox(0,0)[r]{$y\ $}}
          \put(0,100){\makebox(0,0)[r]{$x\ $}}
          \put(12.5,57.5){\makebox(0,0)[t]{$\ t$}}
          \put(28,77){\makebox(0,0)[b]{$\ s$}}
          \put(37.5,67.5){\makebox(0,0)[b]{$r$}}
          \put(50,-5){\makebox(0,0)[t]{$\textstyle{0}$}}
          \put(50,25){\makebox(0,0)[l]{$\ v'$}}
          \put(50,47){\makebox(0,0)[t]{$u'\ \ \ $}}
          \put(50,78){\makebox(0,0)[b]{$\ \ \ u$}}
          \put(50,100){\makebox(0,0)[r]{$v\ $}}
          \put(50,130){\makebox(0,0)[b]{$\textstyle{1}$}}
          \put(62.5,57.5){\makebox(0,0)[t]{$r'$}}
          \put(70,50){\makebox(0,0)[t]{$s'\ $}}
          \put(87.5,67.5){\makebox(0,0)[b]{$t'\ $}}
          \put(100,25){\makebox(0,0)[l]{$\ x'$}}
          \put(100,50){\makebox(0,0)[l]{$\ y'$}}
          \put(100,75){\makebox(0,0)[l]{$\ z'$}}
          \put(100,100){\makebox(0,0)[l]{$\ w'$}}


        \end{picture}
      } 
    \end{picture}

\caption{Rose-Wilkinson lattice\label{fig:mccune4}}
\end{figure}
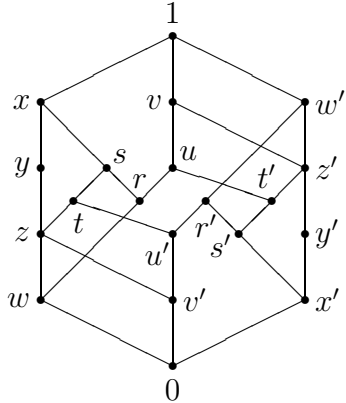

For example, in \cite[Th.\ 3.2]{mpijtp03b} we proved that
equation
\begin{eqnarray}
\hskip-23pt(a\equiv b)\cap((b\equiv c)\cup(a\equiv c))
=((a\equiv b)\cap(b\equiv
c))\cup((a\equiv b)\cap(a\equiv c)),\quad \label{eq:theequation}
\end{eqnarray}
which holds in any OML, does not hold in all WOMLs,
since it fails in the Rose-Wilkinson ortholattice in
Figure \ref{fig:mccune4} which satisfies the WOML
condition Eq.~(\ref{eq:woml2a}).

If we add Eq.~(\ref{eq:theequation}) to the WOML conditions,
we get a family of lattices---let us call it WOMLi---which is
strictly smaller than WOML and strictly larger than OML.
One of its valuations is obviously on the O6 lattice
but not on the Rose-Wilkinson lattice.
In analogy to the way we introduced proper WOMLs in
Section \ref{sec:newwomloml}, we can  define WOMLi-OML
as the class WOMLi$\setminus$OML, each member of which
is a {\em proper} WOMLi.
Now the class WOML contains both the Rose-Wilkinson and O6
lattices. The class WOMLi-OML will contain O6 but not the
Rose-Wilkinson lattice. The class OML will contain neither
Rose-Wilkinson nor O6.  A slight modification of the proof
of Section \ref{sec:compl-ql} (by replacing WOML with
WOMLi) shows that quantum logic is complete for WOMLi-OML, and it is
also complete for WOMLi itself as follows from the completeness
proofs of quantum logic for WOML given in
\cite{mpcommp99,pm-ql-l-hql1}.

Alternatively, we can obtain a hierarchy of classes of models
for quantum logic by adding conditions to the equations
determining the class WOML.
Rather than restricting WOML by subtracting OML from it (to
obtain WOML-OML), we restrict WOML by adding new conditions
(stronger than the WOML law but weaker than
the orthomodular law) to its defining equations to obtain smaller
equational varieties, in between OML and WOML.
We obtain the analogous hierarchy for classical logic by
substituting ``WDOL'' for ``WOML,'' ``BA'' for ``OML,'' and
``distributive'' for ``orthomodular.''
For instance, if we start with WOML, we can choose any model from it
we wish: O6, Rose-Wilkinson, Beran 7b \cite[Fig.\ 7b]{beran},
or any other WOML lattice. When we add the condition
(\ref{eq:theequation}) we can no longer use, e.g., the Rose-Wilkinson
lattice/valuation. When we add the orthomodular law, we can no longer
use O6 or Rose-Wilkinson or Beran 7b valuations.
Thus by adding conditions to the definitions of WOML and WDOL,
we change values (valuations) of logical propositions and we
call this {\em valuation non-monotonicity}. More formally:

\begin{Theorem}\label{th:non-mon}
Quantum (classical) logic is sound and complete with
respect to either the\/ {\rm WOML (WDOL)} or\/ the {\rm OML (BA)}
model families or any model family which is in between\/
{\rm WOML (WDOL)} and\/ {\rm OML (BA)} (such as\/ {\rm WOMLi} above).
Particular {\rm WOML, WDOL, OML, BA, WOMLi} lattices
represent valuation sets for logical propositions.
By adding conditions to Definitions 2.1 and 2.8 {\rm (WOML)}, 2.1. and
(68) {\rm (WOMLi)}, 2.1 and 3.1 {\rm (WDOL)}, etc. we change the sets
of valuations that can be ascribed to propositions. This property
of logical propositions getting new sets of values,
when we add new conditions to the original
definition of lattices to model our logic with, we
call {\em valuation-nonmonotonicity}
\end{Theorem}

{\parindent=0pt{\em Proof. }
The soundness and completeness proofs for WOML and WDOL are given by
Theorems 29 \&\ 39 and 30 \&\ 47 of \cite{pm-ql-l-hql1} (or by Theorems
3.1 \&\ 3.29 and 4.3 \&\ 4.11 of \cite{mpcommp99}),
respectively. The soundness and completeness proofs for OML and BA
are well known. See, e.g., \cite{kalmb74} and  \cite{hilb-ack-book}.
Soundness and completeness proofs for any lattice in between WOML and OML
and in between WDOL and BA follow from the respective proofs for
WOML and OML. For the soundness part of the proof, this is because any
such WOMLj or WDOLj (${\rm j}=1,2,\dots$) is a WOML or WDOL, respectively.
We can obtain a proof that quantum (classical) logic is complete
for WOMLj (WDOLj) by rewriting the completeness proof
of Section \ref{sec:compl-ql} (\ref{sec:compl-cl}) so that the
set of mappings to O6 that refines the equivalence relations is
replaced by a set of mappings to a lattice that satisfies WOMLj
(WDOLj) but violates WOMLj+1 (WDOLj+1), e.g., the Rose-Wilkinson
lattice for WOMLj = WOML and WOMLj+1 = WOMLi.
The part of the proof that refers to
adding conditions is obvious from the very definitions of
WOML, WOMLj, OML, WDOL, WDOLj, and BA.
$\phantom .$\hfill$\blacksquare$}

We stress here that we cannot mix up the two alternative
ways of choosing valuations (restricting classes and
forming set differences vs.\ valuation-nonmonotonicity),
because if we added, e.g., the conditions defining OML (BA)
to WOML-OML (WDOL-BA), we would simply get
empty sets.

\section{Completeness for Smaller Model Subclasses}
\label{sec:compl-smaller}

The reader familiar with the authors' earlier completeness proofs in
\cite{mpcommp99} will notice that the new proofs here, in Sections
\ref{sec:compl-ql} and \ref{sec:compl-cl}, are identical except for the
replacement of WOML (WDOL) with WOML-OML (WDOL-BA) in certain places.
This yields a stronger result for each logic ($\mathcal{QL}$ and
$\mathcal{CL}$), i.e., each is complete for a smaller class of models.
If a logic is complete for a class of models, it obviously continues to
be complete if more models for the logic are added to that class.  Thus
the earlier completeness results follow immediately from the new
ones, since WOML is obtained from WOML-OML by adding back the OML models
for $\mathcal{QL}$ (and analogously WDOL for $\mathcal{CL}$).

The key idea that allowed us to exclude OML from WOML in the
$\mathcal{QL}$ completeness proof was refinement of the equivalence
relation in Theorem \ref{th:congruence-nonoml} with the set of mappings
${\mathcal{O}}{\rm 6}$.  This resulted in smaller equivalence classes,
allowing us to construct a Lindenbaum algebra that violated the
orthomodular law and is thus a proper WOML.

In fact, the ${\mathcal{O}}{\rm 6}$ ``trick'' is not limited to the use
of lattice O6.  We can rewrite the completeness proof for e.g.\
$\mathcal{QL}$ using any lattice that is a proper WOML (a WOML but not
an OML) in place of O6.  This will result in a completeness proof for a
different class of models that can be an even smaller subclass of WOML.

For example, the Rose-Wilkinson lattice of Figure \ref{fig:mccune4} is a
proper WOML.  If we use it in place of O6, an analogous completeness
proof shows that $\mathcal{QL}$ is complete for the class
WOML$\setminus$WOMLi, which is strictly smaller than WOML-OML.  Since
WOML$\setminus$WOMLi doesn't include O6, this shows that
$\mathcal{QL}$ is complete for a class of models that is not only
unrelated to OMLs but is even unrelated to the ``natural'' OML
counterexample O6, which up to now has served as our prototypical WOML
example.

As mentioned earlier, for classical logic $\mathcal{CL}$, we have an
even stronger completeness result that it is complete for {\em single}
WDOL lattices, not just classes of them.  For example, it turns out that
the Rose-Wilkinson lattice is also a proper WDOL (as well as a proper
WOML).  Thus the Rose-Wilkinson lattice, by itself, provides a model for
which classical logic is sound and complete, showing that the hexagon O6
is not the only ``exotic'' non-Boolean lattice model for $\mathcal{CL}$.

\section{Conclusion}
\label{sec:concl}

The main result we obtained in the previous sections is
that logics can be modelled by disjoint classes of
different ortholattices. Classical logic can be modelled
by non-distributive lattices and quantum logic
by non-orthomodular lattices. These lattices
represent different disjoint valuation sets, where
the valuation is a mapping from propositions to a lattice.
Thus by adding conditions (axioms) to
the original definition of an ortholattice
we determine classes of lattices that in turn
determine valuations that one can ascribe to logical
propositions. We call the latter property of logical
propositions {\em valuation-nonmonotonicity} (see
Theorem \ref{th:non-mon}). But by considering disjoint
classes of lattices we can further restrict
valuations we want to use. This can be done as follows.

We considered varieties of classical non-distributive
weakly distributive lattice (WDOL, see Definition
\ref{th:wdol2}) models of classical propositional logic
and non-orthomodular weakly orthomodular lattice (WOML,
Definition \ref{def:oml2}) models of quantum quantum
propositional logics and proved their soundness and
completeness for those models
(see Theorems \ref{th:soundness}, \ref{th:c-soundness},
\ref{th:completeness-woml}, and \ref{th:completeness-wdol-c})

In particular, we considered subclasses of  WDOL and WOML that
do not contain Boolean algebras (BAs, Definition \ref{def:ba})
and orthomodular lattices (OMLs, Definition \ref{def:oml2}),
respectively, while in Sections \ref{sec:nonmonoton} and
\ref{sec:compl-smaller} we also considered a possibly infinite
sequence of subclasses of  WDOL and WOML that do not contain
lattices WDOLi and WOMLi, respectively, which in turn properly
contain BA and OML, and for all of which we have proved the
soundness and completeness.  We denoted these classes (varieties
of WDOL and WMOL) as WDOL-BA, WOML-OML, WDOL-WDOLi, and
WMOL-WOMLi. The valuations of WOML-OML and OML, of WDOL-BA
and BA, of WODL-WODLi and WODLi, of
WOML-WOMLi and WOMLi [Eq.\ (\ref{eq:theequation})], and of
WOMLi-OML and OML do not overlap. For instance, valuations from
WDOL-BA cannot be numeric (\{0,1\} or \{{\1TRUE,FALSE}\}) at
all since it does not contain the two-valued Boolean algebra.

At the level of logical gates, classical or
quantum, with today's technology for computers and artificial
intelligence, we can use only
bits and qubits, respectively, i.e., only valuations
corresponding to \{0,1\} BA and OML, respectively. And when
we talk about logics today, we take for
granted that they have the latter valuation---\{{\1TRUE,FALSE}\}
in the case of classical logic and Hasse (Greechie) diagrams
in the case of quantum logic \cite{bdm-ndm-mp-1}. This is because
a valuation is all we use to implement a logic. In its final
application, we do
not use a logic as given by its axioms and rules of
inferences but instead as given by its models. Actually,
logics given only by their axioms and rules of inferences
(in Sections \ref{subsec:q-logic} and \ref{subsec:c-logic}),
i.e., without any models and any valuations, cannot be
implemented in any hardware at all.

It would be interesting to investigate how
other valuations, i.e., various ortholattices,
might be implemented in complex circuits. That would provide
us with the possibility of controlling essentially different
algebraic structures (logical models) implemented into radically
different hardware (logic circuits consisting of logic gates)
by the same logic as defined by its axioms and rules
of inference.

\bigskip
{\parindent=0pt{\bf Acknowledgement}
Supported by the Ministry of
Science, Education, and Sport of Croatia through the project
No.\ 082-0982562-3160.}

\end{document}